\documentclass[a4paper,fleqn]{cas-dc}

\usepackage[numbers,sort&compress]{natbib}
\def\tsc#1{\csdef{#1}{\textsc{\lowercase{#1}}\xspace}}
\tsc{WGM}
\tsc{QE}
\tsc{EP}
\tsc{PMS}
\tsc{BEC}
\tsc{DE}
\usepackage{moreverb,url}
\usepackage{svg}
\usepackage{amsmath}
\usepackage{supertabular}
\usepackage{subfig}

\usepackage{caption}
\usepackage{placeins}
\usepackage{bm}
\usepackage{soul}

\begin{document}
\let\WriteBookmarks\relax
\def\floatpagepagefraction{1}
\def\textpagefraction{.001}

\shorttitle{Realizability-Constrained Machine Learning for Turbulence Closures in Wake Flows}

\shortauthors{Ansari et~al.}


\title [mode = title]{Realizability-Constrained Machine Learning for Turbulence Closures in Wake Flows}




%

\author[1]{Talib Ansari}
\credit{Conceptualization, Data curation, Formal analysis, Visualization, Software, Writing - original draft, Writing - review \& editing, Methodology, Validation}

\author[1]{Priyank H. Mehta}
\credit{Software, Conceptualization, Writing - review \& editing, Formal analysis}

\author[1]{Harshal D. Akolekar}[type=editor,
                        auid=000,bioid=1,
                        orcid=0000-0002-3178-2987]

\cormark[1]


\ead{harshal.akolekar@iitj.ac.in}


\credit{ Conceptualization, Formal analysis, Software, Supervision, Writing - original draft, Writing - review \& editing, Funding}

\affiliation[1]{organization={Indian Institute of Technology Jodhpur},
    city={Jodhpur},
    postcode={342030}, 
    country={India}}




\begin{abstract}
Computational fluid dynamics (CFD)-driven machine learning frameworks based on symbolic regression offer a promising pathway for turbulence model discovery, but are often hindered by numerical instability, residual stagnation, and non-physical model behavior during training. In particular, realizability, which is rarely enforced explicitly during model development, remains a critical yet overlooked requirement, especially for accurate wake prediction. In this work, a residual- and realizability-filtered CFD-driven framework is proposed to enhance both efficiency and robustness within a gene expression programming (GEP) paradigm. The method integrates two residual-based filtering criteria along with a barycentric-map-based realizability constraint directly into the CFD solution loop, enabling early identification and rejection of unstable and non-realizable candidate models. This reduces unnecessary computational effort while guiding the search toward physically admissible solutions. The proposed approach achieves a 42.3\% reduction in computational cost relative to the baseline CFD-driven GEP framework and reduces non-realizable models at convergence from 58.4\% to 1.7\%. The framework is trained on a canonical cylinder wake. The resulting models enhance mean wake prediction and remain realizable across training and test cases, with robust generalization to diverse geometries and operating conditions, including a rectangular cylinder, an airfoil, and an axisymmetric body. The study further provides insights into realizable model statistics, coefficient trends, and conditions governing physically consistent wake behavior. These results demonstrate that incorporating realizability and stability constraints within CFD-driven learning enables efficient and physically consistent turbulence model discovery, offering a scalable pathway toward reliable data-driven closure development.
\end{abstract}





\begin{keywords}
Machine learning \sep Turbulence modelling \sep Realizability \sep Wake Flow \sep Gene Expression Programming
\end{keywords}

\maketitle

\section{Introduction}
Reynolds-averaged Navier–Stokes (RANS) simulations remain the mainstay of iterative industrial design workflows due to their relatively low computational cost and ease of use \citep{Sandberg2022}. However, their predictive accuracy is often limited when compared to high-fidelity simulations, primarily owing to the reliance on the Boussinesq approximation \citep{Schmitt2007} as a closure hypothesis in linear eddy-viscosity turbulence models. This approximation enforces a linear relationship between the Reynolds stress anisotropy and the mean strain-rate tensor, which is known to be inadequate for complex turbulent flows involving strong anisotropy, curvature, separation, or non-equilibrium effects \citep{Leschziner2015,Sandberg2022}.

In recent years, machine learning–based approaches have emerged as a promising pathway to improve the fidelity of RANS predictions, particularly by non-linearizing the Boussinesq approximation  \citep{Pope1975,Ling2016a,Weatheritt2016,Kaandorp2020}. Among the various data-driven methodologies explored, symbolic regression–based approaches have gained significant traction, as they yield explicit, physics-interpretable closure expressions rather than black-box models  \citep{Weatheritt2016,Schmelzer2020,max2025}. In this context, gene expression programming (GEP)–based symbolic regression has been successfully applied across a wide spectrum of flow configurations. These include the development of explicit algebraic Reynolds stress models (EARSM) for canonical benchmark cases such as periodic hills \citep{Weatheritt2016} and duct flows \citep{Weatheritt2017a}, as well as more complex industrially relevant flows, including low-pressure turbines \citep{Akolekar2019,Akolekar2019a,Akolekar2021a,Gu2025}, high-pressure turbines  \citep{Weatheritt2017,Zhao2020,Akolekar2019isabe}, combustion \citep{schoepplein2018}, and cooling flows \citep{Haghiri2020,Sandberg2018a}.

Broadly, two principal approaches exist for deploying GEP in turbulence modeling. The first is the so-called frozen approach \citep{parneix1998procedure,Weatheritt2016,Akolekar2019a}, wherein GEP is trained offline using large, precomputed high-fidelity datasets and subsequently deployed in a static manner within RANS solvers. While computationally efficient, this approach does not explicitly account for solver feedback or numerical stability during the training process. The second approach is the computational fluid dynamics (CFD)-driven machine learning framework \citep{Zhao2020}, which tightly integrates the GEP algorithm with a CFD solver in a closed-loop manner. This framework enables candidate models to be evaluated in real time as part of the solver execution, thereby promoting the development of numerically robust, solver-consistent, and dynamically adaptive closure models.

Several extensions of the CFD-driven framework have been proposed to further enhance model capability and robustness. Multi-objective and multi-expression CFD-driven formulations \citep{waschkowski2022multi} enable simultaneous optimization against multiple physical targets, facilitating the development of coupled turbulence–heat-flux models \citep{xu2022towards} as well as transition–turbulence closures. \citep{Akolekar2022} To improve generalization and robustness across flow regimes, multi-case training strategies have been introduced, wherein models are trained concurrently on multiple flow configurations \citep{Fang2023TowardTraining,fang2024data}. More recently, transformer-based methodologies have been incorporated into the GEP model discovery process to guide and structure the search for improved closure expressions\citep{max2025,Fang2026}.

However, despite their demonstrated accuracy improvements, CFD-driven machine-learning frameworks face an important and often overlooked limitation: numerical instability during model discovery. In CFD-driven training, thousands of candidate closures are sequentially embedded into a RANS solver and iterated toward convergence to evaluate their fitness. A non-negligible fraction of these learned models leads to solver divergence, residual stagnation, or unphysical growth of turbulence quantities when coupled with the governing equations. Such unstable behavior is often only detected after several thousand iterations, by which point substantial computational resources have already been expended. As a result, CFD-driven training pipelines can become computationally inefficient, with a significant portion of total wall-clock time devoted to simulations that ultimately do not yield usable turbulence models. None of the existing CFD-driven approaches selectively filter out unstable models during the training process. 

In addition to numerical stability, realizability of the learned turbulence closures represents another critical yet largely unaddressed limitation of existing CFD-driven machine-learning frameworks. Realizability constraints ensure that modeled Reynolds stresses remain physically admissible, for instance by enforcing positivity of turbulent kinetic energy, bounded anisotropy, and compliance with fundamental tensor invariants \citep{Lumley1970,Banerjee2007,Leschziner2015}. In current CFD-in-the-loop training paradigms, candidate closure models are primarily evaluated based on their impact on the mean-flow behavior, with little to no explicit enforcement of realizability conditions during model discovery. As a result, learned closures may yield Reynolds stress states that violate physical constraints, even when the corresponding simulations exhibit improved agreement with reference data \citep{Akolekar2019thesis,Fang2023TowardTraining}. Such violations can compromise model robustness, lead to non-physical flow predictions under extrapolative conditions, and limit the general applicability of the resulting closures. This underscores the need for incorporating realizability-aware constraints or filters directly within CFD-driven learning pipelines to ensure that improvements in accuracy are accompanied by physically consistent stress representations. 
Enforcing realizability within such training frameworks is, however, inherently challenging, as the physical admissibility of the learned closure is not determined solely by the functional form of the model, but by its interaction with the post-processed flow field obtained after solver convergence \citep{Wang2017,Wolff2025}. In particular, realizability violations often emerge only after the reconstructed Reynolds stresses are assembled, normalized, and evaluated through eigenvalue- or invariant-based criteria, which are typically applied as \textit{a posteriori} checks rather than being embedded within the learning process. 
This separation between model generation and realizability evaluation further limits the ability of existing CFD-driven frameworks to actively steer the learning process toward physically admissible stress states.

The present work addresses this challenge by introducing a stability- and realizability-aware supervisory layer designed to improve the efficiency and robustness of CFD-driven turbulence model discovery without compromising predictive accuracy. Specifically, a residual- and realizability-filtered CFD-driven algorithm is developed within an established GEP–based framework. The approach integrates two residual-based stopping criteria followed by a realizability-based filter directly into the RANS solution process. The first two stages monitor the absolute magnitude of solver residuals and their reduction over a prescribed iteration window, enabling early detection of stagnation or divergence. Candidate models that fail these checks are terminated prematurely, preventing unnecessary computational expenditure. The third stage enforces realizability through a barycentric-map-based constraint, providing a complementary physical filter that enhances the robustness of the discovered closures.

To demonstrate the framework, turbulent flow past a circular cylinder \citep{parnaudeau2008cylinder3900} is used as the training configuration. This canonical bluff-body flow exhibits strong separation and complex wake dynamics, making it a challenging benchmark for turbulence modeling and particularly suitable for data-driven model discovery. Furthermore, the circular cylinder wake has not previously been employed as a training configuration within CFD-driven GEP-based turbulence model development. In order to illustrate their robustness, the models obtained from the proposed framework are subsequently evaluated on several additional geometries, including a 5:1 rectangular cylinder \citep{mannini2019benchmarkRectangularCylinder}, the National Advisory Committee for Aeronautics (NACA 0012) airfoil \citep{lin2013wallResolvedLESNACA0012}, and the axisymmetric Defense Advanced Research Projects Agency (DARPA) Suboff \citep{jimenez2010intermediateWake}. These tests demonstrate the ability of the framework to discover realizable and robust turbulence closures that improve wake predictions across diverse flow configurations while enabling exploration of a broader model space at reduced computational cost.

\FloatBarrier
\section{\label{sec:level1}Methodology \protect\\}
This section describes the numerical setup for cylinder flows and three test cases, EARSM formulation and the filtered CFD-driven machine learning algorithm.
\subsection{\label{sec:level2}Numerical Setup}
\subsubsection{Computational Domain for Cylinder}

\begin{figure}
\centering
\includegraphics[width=0.4\textwidth]{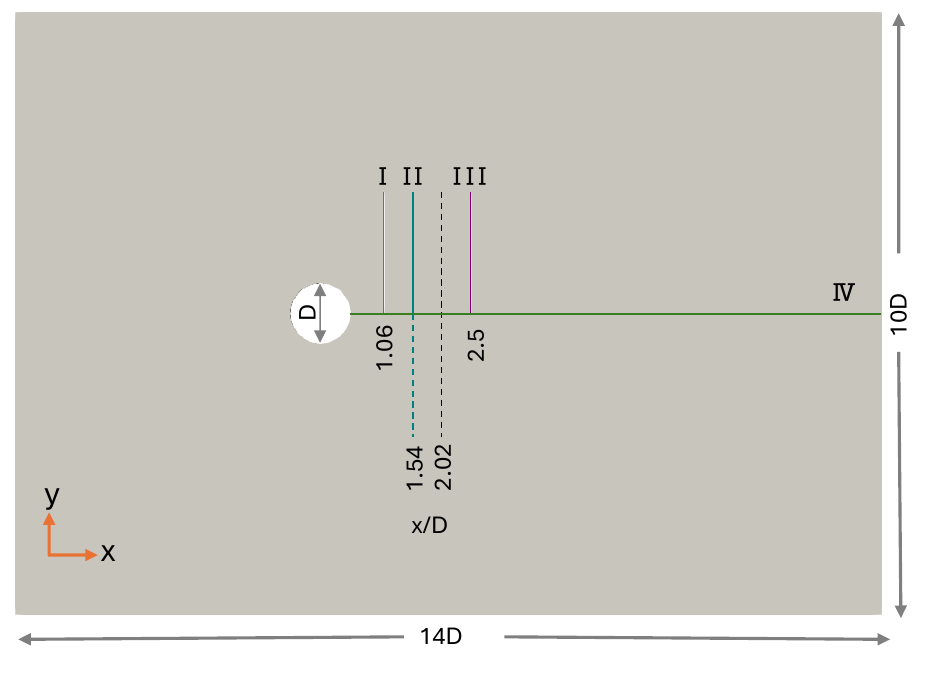}
\caption{\label{fig:cylinder_schematic} The computational domain for a cylinder at Re=3900. }
\end{figure}

The computational domain, as shown in Fig. \ref{fig:cylinder_schematic}, extends 14D in the streamwise direction and 10D in the transverse direction, with a downstream wake development region of 9D from the cylinder center. 
The Reynolds number based on the diameter (D=0.12m) was Re=3900. 
Boundary conditions were prescribed as follows: a uniform x-velocity 
was imposed at the inlet, with a freestream turbulence intensity $Tu_{\infty}=0.2\%$ and inlet length scale $l_T /D = 0.07$. At the outlet, zero-gradient conditions were applied for velocity and turbulence variables, while pressure was fixed to provide a reference level. A no-slip boundary condition was enforced on the cylinder surface.  Freestream boundary conditions were imposed at the top and bottom boundaries. A few wake measurement locations are also shown in Fig. \ref{fig:cylinder_schematic}. 

\subsubsection{\label{sec:level2a}Grid Convergence \& Validation}
A structured mesh was generated using the OpenFOAM BlockMesh utility. A multi-block strategy was adopted to maintain mesh quality and numerical robustness. An O-grid topology was employed in the vicinity of the cylinder to accurately capture boundary-layer gradients and maintain near-orthogonal cells around the curved surface. The remainder of the domain was discretized using an H-grid topology to preserve alignment with the main flow direction and reduce numerical diffusion in the wake.

\begin{figure}
\includegraphics[width=0.45\textwidth]{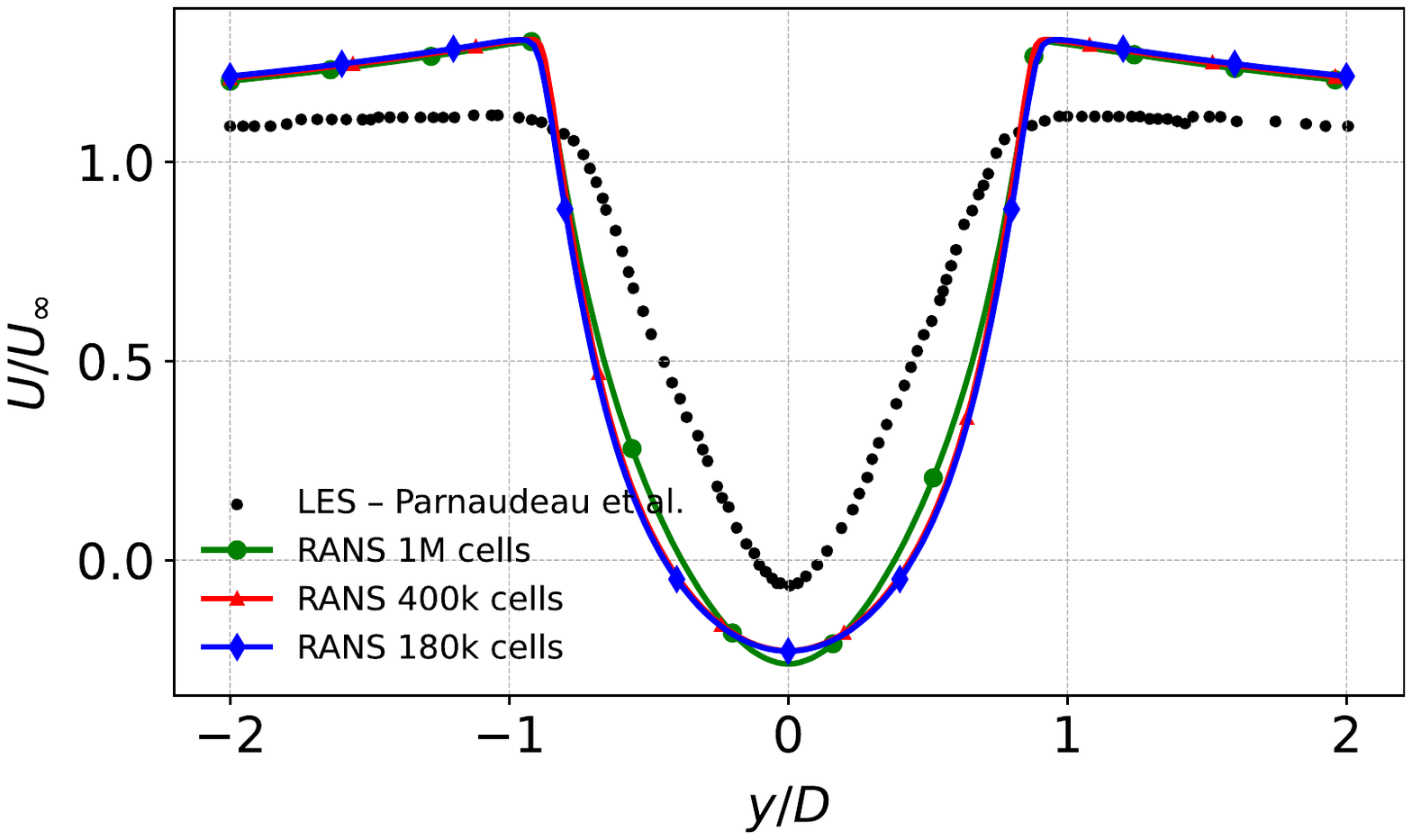}
\caption{\label{grid_convergence} Streamwise velocity profile for grid convergence at $x/D = 2.02$.}
\end{figure}

Special attention was devoted to wall-normal grading to ensure that the first cell height satisfies $y^{+} < 1$. 25-30 points were present in the boundary layer.  
Three mesh levels were investigated---180k, 400k, and 1 million,  
and grid independence was verified by comparing solutions in the wake region. Figure \ref{grid_convergence} shows the wake profiles at $x/D=2.02$ for three different grid levels, with respect to the experimentally-validated large eddy simulations (LES) of Parnaudeau et al. \citep{parnaudeau2008cylinder3900}.
The differences between the three grid levels are very minimal and within 1\%. Hence, the 180k mesh was selected for further analysis. The RANS calculations also accurately captured the pressure distribution on the blade surface, similar to the LES. 

\subsubsection{\label{sec:level2b}Computational Domains for Test Cases}

\begin{figure}
\centering
\includegraphics[width=0.4\textwidth]{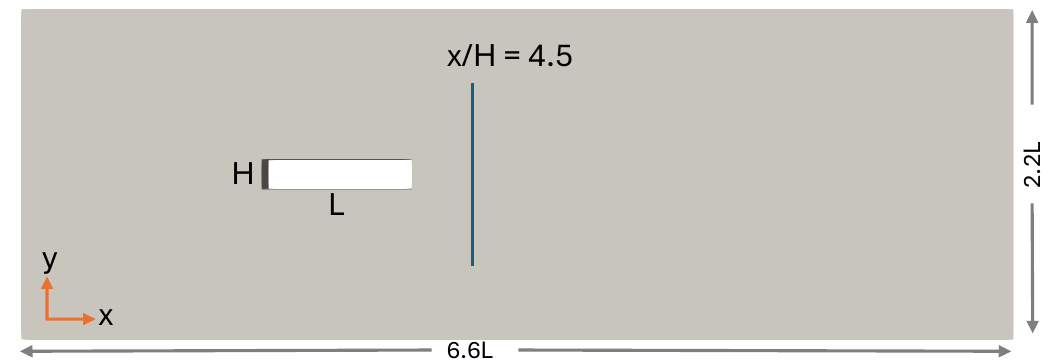}
\caption{\label{fig:rect_schematic} The Computational domain of the rectangular 5:1 cylinder.}
\end{figure}

A total of three test cases were are used in this study and their computational domains are shown. Figure \ref{fig:rect_schematic} depicts the CFD setup for a rectangular 5:1 cylinder, a canonical bluff-body configuration widely studied in wind engineering applications for understanding wake dynamics, separation behavior, and aerodynamic loading \citep{mannini2019benchmarkRectangularCylinder}. The boundary conditions are consistent with the cylinder case, with a prescribed inlet velocity and standard outlet and freestream and zero-gradient conditions to achieve a Reynolds number of Re=44,900. 

\begin{figure}
\centering
\includegraphics[width=0.4\textwidth]{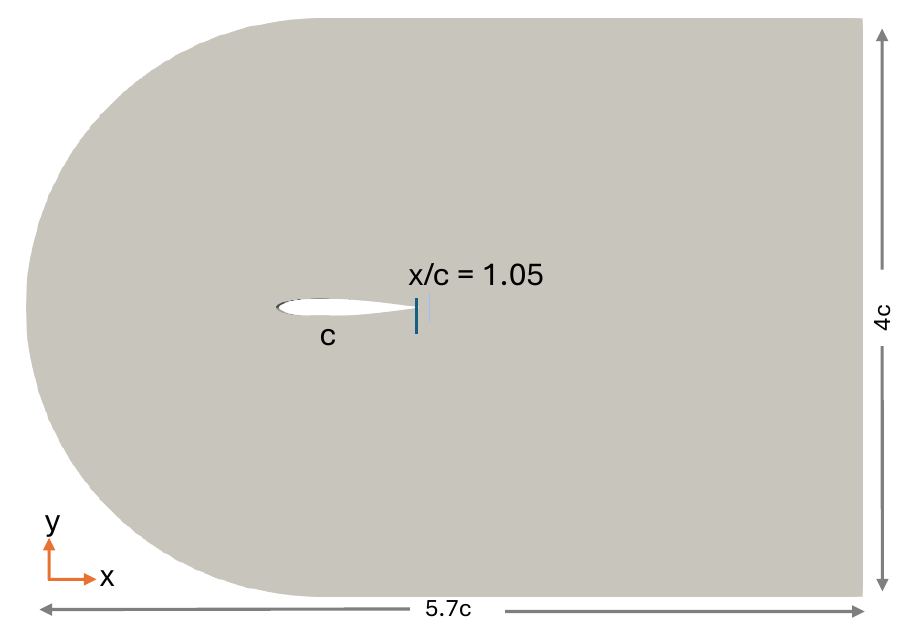}
\caption{\label{fig:naca_geometry} The computational domain of the NACA 0012.}
\end{figure}

The computational domain of a NACA 0012 airfoil, as shown in Fig. \ref{fig:naca_geometry} extends sufficiently in all directions to minimize boundary interference. A uniform inflow velocity is prescribed at the inlet to achieve a Reynolds number of Re=2x10$^5$ based on the chord length (\textit{c}). A no-slip boundary condition is imposed on the airfoil surface, while zero-gradient conditions are applied at the outlet. Freestream boundary conditions are enforced at the far-field boundaries.  

\begin{figure}
\centering
\includegraphics[width=0.95 \columnwidth]{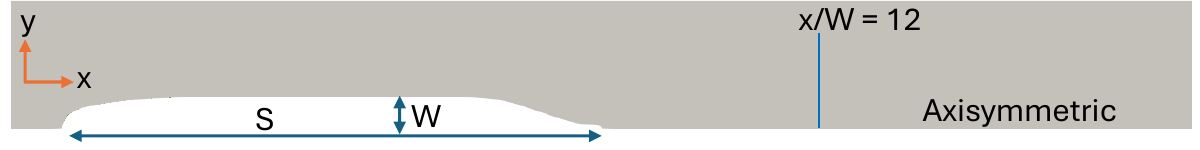}
\caption{\label{fig:darpa_schematic} Partial computational domain of the DARPA Suboff.}
\end{figure}

The partial computational domain of the DARPA Suboff, an axisymmetric body of revolution, is shown in Fig.\ref{fig:darpa_schematic}.  Simulations are performed at $\mathrm{Re}=1.1\times10^{6}$ based on the body length $S$, which tests the model performance in high-Reynolds-number conditions with mild separation and complex wake recovery. The Suboff geometry consists of a smooth forebody, cylindrical mid-section, and an afterbody with gradual tapering, ensuring predominantly attached flow along the surface and a well-defined axisymmetric wake. 
The computational domain is constructed with sufficient upstream (3S), radial, and downstream extent (7S) to eliminate boundary interference effects. A uniform inflow velocity is prescribed at the inlet, while zero-gradient conditions are applied at the outlet. A no-slip boundary condition is enforced along the body surface with complete near-wall resolution. Axisymmetric boundary conditions are applied along the centerline, and far-field boundaries are treated using symmetry conditions to approximate free-stream behavior. 

Grid convergence studies were conducted for all test cases and the most optimum mesh was used. Additionally, wake measurement locations are shown for all the three test cases. 


\subsubsection{\label{sec:level2d}Schemes and Solvers }

Steady incompressible RANS equations were solved using a finite-volume formulation within OpenFOAM. 
Pressure–velocity coupling was handled using the semi-implicit method for pressure-linked equations  (SIMPLE) algorithm. Second-order accurate spatial discretization schemes were employed for convective and diffusive terms to minimize numerical dissipation in the separated shear layers and wake. Gradients were evaluated using second-order schemes to ensure consistency. Under-relaxation factors were selected to guarantee stable convergence of all transport equations.
The baseline turbulence model employed was the $k\text{-}\omega$ SST model  \citep{menter1994two}, whose transport equations for the turbulent kinetic energy (\textit{k}) and specific dissipation ($\omega$) are:

\begin{equation}
\frac{\partial k}{\partial t} 
+ \bm{U}_j \frac{\partial k}{\partial x_j} 
= P_k  
- \beta^{*} \, \omega\, k 
+ \frac{\partial}{\partial x_j} 
  \left[ \left( \nu + \sigma_k \nu_t \right) 
  \frac{\partial k}{\partial x_j} \right],
\label{eq:k_omega_k}
\end{equation}

\begin{equation}
\frac{\partial  \omega}{\partial t} 
+ \bm{U}_j \frac{\partial \omega}{\partial x_j} 
= \frac{\gamma P_k}{\nu_t}
- \beta \, \omega^{2} 
+ \frac{\partial}{\partial x_j} 
  \left[ \left( \nu + \sigma_\omega \nu_t \right) 
  \frac{\partial \omega}{\partial x_j} \right]
+ CD_{k\omega},
\label{eq:k_omega_omega}
\end{equation}

where $P_k = \bm \tau_{ij}\frac{\partial \bm U_i}{\partial x_j}$ which depends on the Reynolds stresses ($\tau_{ij}$). The divergence term of the Reynolds stresses is also present in the RANS momentum equations.

\subsection{Explicit Algebraic Reynolds Stress Models}
The limitations of linear eddy-viscosity models for separated bluff-body flows in predicting the wake flow accurately, primarily stems from their reliance on the Boussinesq approximation, which assumes an isotropic turbulent viscosity relating Reynolds stresses linearly to the mean strain-rate tensor (i.e. a linear anisotropy tensor). In RANS modeling, the Reynolds stress tensor ($\bm{\tau}_{ij}$) is most commonly approximated using the Boussinesq hypothesis as: 
 \begin{equation}
\underbrace{\bm \tau_{ij}}_{\text{Reynolds Stress}} = \underbrace{\frac{2}{3} {\rho k\bm \delta_{{ij}}}}_{\text{isotropy tensor}}+ \underbrace{-2\mu_T \bm S_{ij}'.}_{\text{anisotropy tensor} \hspace{1mm} (\bm a_{ij}^B)}
\label{eqn:boussinesq}
\end{equation}
EARSMs provide a computationally efficient method to enhance the accuracy of the anisotropy tensor ($\bm a_{ij}^B$). These models incorporate additional terms, referred to as tensor bases ($\bm T_{ij}^{(n)}$), which are functions of non-dimensional strain  ($s_{ij} = \tau\bm S_{ij}'$) and rotation-rate ($w_{ij} = \tau\bm \Omega_{ij}$) tensors; where $\tau$ is the turbulent time scale denoted by $(1/\omega)$.  For two-dimensional mean flows, the reduced tensor basis consists of
\begin{equation}
\begin{aligned}
\bm T_{ij}^{(1)} &= \bm s_{ij},
\qquad
\bm T_{ij}^{(2)} = \bm s_{ik}\bm w_{kj} - \bm w_{ik}\bm s_{kj}, \\
\bm T_{ij}^{(3)} &=
\bm s_{ik}\bm s_{kj}
- \frac{1}{3}\bm \delta_{ij}\,\bm s_{mn}\bm s_{nm}, \\
\bm T_{ij}^{(4)} &=
\bm w_{ik}\bm w_{kj}
- \frac{1}{3}\bm \delta_{ij}\,\bm w_{mn}\bm w_{nm}.
\end{aligned}
\end{equation}

By introducing these terms, EARSMs effectively non-linearize the anisotropic relationship, enabling a more precise representation of complex turbulent flows. Equation \eqref{eqn:EARSMd} outlines the EARSM formulation:
\begin{equation}
\bm a_{ij}^{EARSM}=\bm a_{ij}^B  + \hspace{1mm}\sum_{n=1}^{4} \zeta^{(n)}(I_1,I_2)\, \bm T_{ij}^{(n)},
\label{eqn:EARSMd}
\end{equation}  
where the scalar coefficients $\zeta_n$ are functions of the invariants of the non-dimensionalized strain and rotation rate tensor as: 
\begin{equation}
I_1 = \bm s_{mn}\bm s_{nm}; 
\qquad 
I_2 = \bm w_{mn} \bm w_{nm}.
\end{equation}

The scalar coefficient $\zeta_n$ are determined symbolically using GEP. The tensor basis and invariants follow the integrity basis framework of Pope \citep{Pope1975}, thereby ensuring Galilean invariance.

\subsection{Filtered CFD-Driven GEP Algorithm}

\begin{figure*}[!t]
\centering
\includegraphics[width=\textwidth]{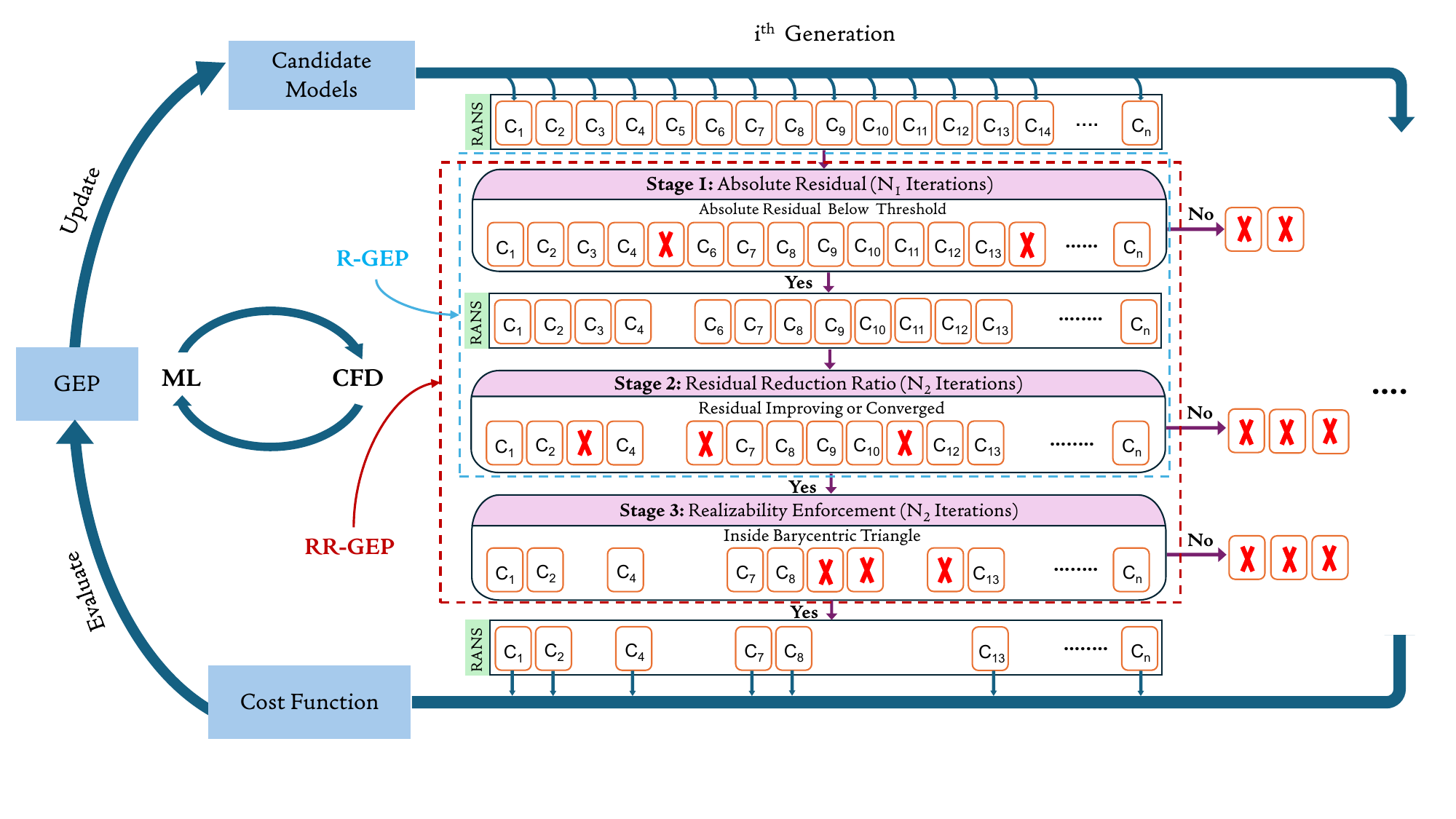}
\caption{\label{fig:filtered_schematic} Filtered residual and realizability GEP-based CFD-driven framework.}
\end{figure*}

The development of CFD-driven turbulence models using evolutionary algorithms is fundamentally constrained by computational cost. In a conventional CFD-driven GEP framework (baseline-GEP or B-GEP), each candidate EARSM must be fully evaluated within a RANS solver to assess predictive accuracy \citep{Zhao2020}. A significant fraction of candidates, however, lead to numerical divergence, residual stagnation, or unphysical stress states, yet still consume thousands of solver iterations before being discarded. With thousands of candidates evaluated, this makes the optimization prohibitively expensive. 
The challenge is amplified in separated flows, where small changes in the stress–strain relationship can strongly perturb separation, wake diffusion, and turbulence production, often destabilizing the solver before meaningful flow structures emerge. Additionally, some candidates produce deceptively favorable objective values despite unstable behavior, arising from partially converged solutions and manifesting as non-smooth or oscillatory wake profile\citep{Ansari2025CFD}. In conventional frameworks, such cases are not identified early, leading to substantial computational effort spent on unstable or physically unreliable models.

To address this limitation, the present work introduces a filtered CFD-driven machine learning algorithm that embeds convergence-aware screening directly within the evolutionary loop. The key innovation is the integration of early-stage residual and realizability-based filtering mechanisms that terminate unproductive candidates before full convergence is attempted. Rather than treating solver stability as \textit{a posteriori} information, it is incorporated as an active decision variable within the learning process.

The overall workflow of the filtered CFD-driven framework is illustrated schematically in Fig. \ref{fig:filtered_schematic}. An initial population of algebraic expressions ($\bm a_{ij}^{EARSM}$) representing the scalar coefficients of the tensor basis is generated using GEP encoding. Each candidate model ($C_n$) is inserted into the RANS solver. The solver is then advanced for a predefined number of iterations to assess convergence behavior. After $N_1$ iterations, the filtering mechanism is activated. Instead of allowing every candidate to iterate to full residual convergence, convergence behavior is evaluated at an intermediate iteration count. The filtering procedure consists of three sequential checks designed to identify unstable or non-realizable solutions early in the evaluation process. 

\subsubsection{Stage 1: Absolute Residual} 

The first criterion evaluates the final residual magnitude of the governing equations. Let $R^{(q)}(N_1)$ denote the normalized residual of variable $q$ at iteration $N_1$, where $q$ represents pressure, velocity components, $k$, or $\omega$. A candidate model is rejected if
\begin{equation}
\max_{q}\left[ R^{(q)}(N_1) \right] > \epsilon_{1},
\end{equation}
where $\epsilon_{1}$ is a prescribed absolute residual threshold. The absolute residual threshold serves as a practical indicator of numerical instability, with values above $\epsilon_{1} = \mathcal{O}(10^{-1})$ reliably identifying divergent or non-convergent candidate models based on preliminary experiments.
This stage eliminates models that produce immediate divergence or excessively large residual levels, which are likely to diverge after a few iterations.

In incompressible RANS simulations, the maximum residual is typically associated with the pressure equation. As pressure is obtained from a globally coupled Poisson equation enforcing mass conservation, segregated algorithms such as SIMPLE lead to domain-wide coupling, causing pressure residuals to converge more slowly and often exhibit larger or oscillatory behavior compared to momentum and turbulence equations
\citep{patankar1980,versteeg2007,FerzigerPeric2020}.
Some models are typically rejected during this stage (iteration $N_1$) and do not proceed further.

\begin{figure}
\centering
\includegraphics[width=0.35\textwidth]{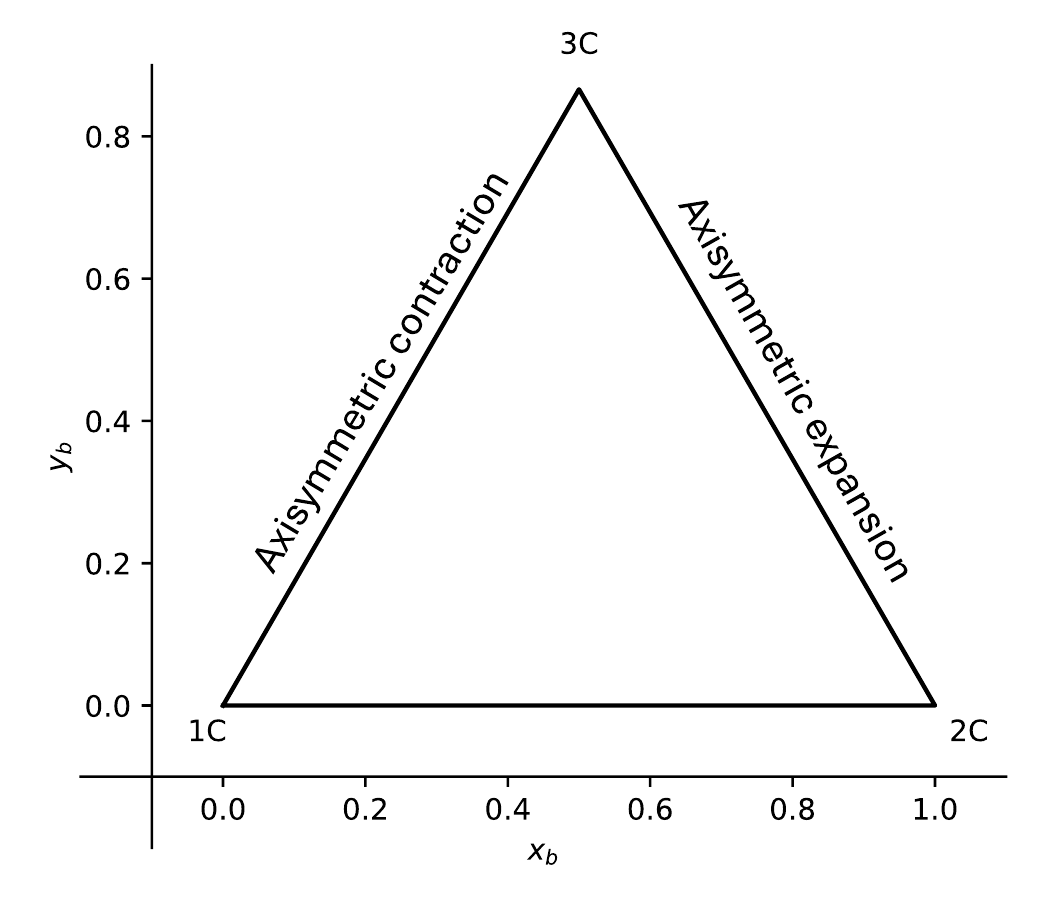}
\caption{\label{fig:triangle} The barycentric map.}
\end{figure}

\subsubsection{Stage 2: Residual Reduction Ratio} Candidates passing the first check proceed to the second filtering stage, which evaluates residual improvement between two iteration levels $N_1$ and $N_2$. The residual reduction ratio is defined as
\begin{equation}
\Gamma =
\frac{
 \max_{q} R^{(q)}(N_1)
}{
\max_{q} R^{(q)}(N_2)
} .
\end{equation}
A candidate clears the second stage only if $\Gamma \ge \gamma_{\min}$ ($10$) or $\max_{q} [ R^{(q)}(N_2)$] < $\epsilon_{2}$  $(10^{-6})$,  where $\gamma_{\min}$ and  $\epsilon_{2}$ represents the minimum acceptable improvement factor (order-of-magnitude reduction) and convergence threshold. This condition ensures that residuals decrease at a sufficient rate and prevents stagnating models from consuming further computational resources. If the solution already satisfies full convergence criteria at $N_2$ (residuals below prescribed final tolerance $\epsilon_{2}$) that is $\max_{q} [R^{(q)}(N_2)]$ < $\epsilon_{2}$, no further residual reduction is required for acceptance. Some more models are rejected at this stage of evaluation. 
The remaining models are further evaluated at the same iteration $N_2$, using a realizability-based constraint. The residual reduction ratio thus provides a solver-independent measure of convergence behavior, as it captures the relative decay of residuals.

\subsubsection{Stage 3: Realizability Enforcement}
Even if residuals decrease satisfactorily, the predicted Reynolds stress tensor may violate realizability constraints. 
This stage of the filtered framework enforces physical admissibility of the modeled Reynolds stresses through an eigenvalue-based barycentric representation of turbulence anisotropy \citep{Banerjee2007} (refer Fig. \ref{fig:triangle}).


Let $\lambda_1 \ge \lambda_2 \ge \lambda_3$ denote the ordered eigenvalues of the normalized anisotropy tensor $\bm a_{ij}^{EARSM}$, subject to the traceless constraint
\begin{equation}
\lambda_1 + \lambda_2 + \lambda_3 = 0.
\end{equation}

Any realizable Reynolds stress state can be expressed as a convex combination of three limiting turbulence componentality states: one-component (1C), two-component (2C), and three-component isotropic (3C) turbulence, and must lie within the Barycentric map. In a Euclidean coordinate system, these limiting states are positioned at the vertices of an equilateral triangle,
$\mathbf{x}_{1c} = (1,0), \quad
\mathbf{x}_{2c} = (0,0), \quad
\mathbf{x}_{3c} = (1/2, \sqrt{3}/2).
$
The barycentric coordinates $(x_B, y_B)$ of a given anisotropy state are obtained as a linear combination of these vertices,
\begin{equation}   
\begin{aligned}
x_B &= C_1 x_{1c} + C_2 x_{2c} + C_3 x_{3c}
     = C_1 + \frac{1}{2}C_3, \\
y_B &= C_1 y_{1c} + C_2 y_{2c} + C_3 y_{3c}
     = \frac{\sqrt{3}}{2} C_3,
\end{aligned}
\end{equation}
where the barycentric weights are defined directly from the eigenvalues as
\begin{equation}
\begin{aligned}
C_{1c} &= \lambda_1 - \lambda_2, \\
C_{2c} &= 2(\lambda_2 - \lambda_3), \\
C_{3c} &= 3\lambda_3 + 1.
\end{aligned}
\end{equation}
To ensure realizability, the barycentric weights must satisfy the convexity constraints
\begin{equation}
    C_i \ge 0; \hspace{4mm} \sum_{i=1}^{3} C_i = 1; \hspace{3mm} i=1,2,3. 
\end{equation}

In the present filtered CFD-driven framework, the barycentric map is enforced as a realizability constraint at iteration $N_2$. The anisotropy tensor is evaluated pointwise across $M$ data-points of the computational domain in the wake region, its eigenvalues are computed and ordered, and the corresponding barycentric weights are determined.

Because the RANS solution has not necessarily reached asymptotic convergence at this stage, residual-level perturbations in the governing equations can lead to small numerical violations of the realizability constraints. To account for these effects, during partially converged iterations, a small realizability tolerance $\varepsilon_3$ is introduced. A candidate closure is not rejected at this stage only if the barycentric weights satisfy
\begin{equation}
    C_i \ge -\varepsilon_3; \hspace{4mm}     \left| \sum_{i=1}^{3} C_i - 1 \right| \le \varepsilon_3; \hspace{3mm} i=1,2,3.
\end{equation}

Thus, the barycentric coordinate vector is required to remain within an $\mathcal{O}(\varepsilon_3)$ neighborhood of the realizable barycentric domain.
The tolerance is scaled using an effective residual defined as
\begin{equation}
R_{\mathrm{eff}} = \max \left(R_P, R_{U}, R_{k}, R_{\omega} \right),
\end{equation}
where $R_P$, $R_{U}$, $R_{k}$, and $R_{\omega}$ denote the normalized residuals of the pressure, momentum, turbulent kinetic energy, and specific dissipation rate equations, respectively. Since the anisotropy tensor depends on velocity gradients and modeled Reynolds stresses, perturbations in these equations propagate directly to the stress tensor; using the maximum residual therefore provides a conservative estimate of the perturbation level in the anisotropy tensor. The realizability tolerance is defined as
\begin{equation}
\varepsilon_3 = \alpha R_{\mathrm{eff}} .
\end{equation}

In steady segregated RANS solvers, 
velocity perturbations are typically one to two orders of magnitude larger than the maximum residual level \citep{FerzigerPeric2020}. Because the barycentric coordinates depend on the eigenvalues of the normalized anisotropy tensor $\bm a_{ij}^{EARSM}$, matrix perturbation theory implies that eigenvalue variations scale with perturbations in $\bm a_{ij}^{EARSM}$ \citep{HornJohnson2013}. Consequently, residual levels $R_{\mathrm{eff}} \sim 10^{-4}$ can induce barycentric coefficient variations of order $10^{-2}$. An amplification factor $\alpha \sim 10^{2}$ is therefore adopted, giving $\varepsilon_3 \sim 10^{-2}$.

The tolerance is applied exclusively at iteration $N_2$ to account for incomplete convergence. Fully converged solutions (iteration $N_3$) must satisfy strict realizability without tolerance. Any violation of the barycentric admissible domain at any of the $M$ locations results in immediate rejection of the candidate closure. 
Some more models are rejected during stage 3. The remaining models continue till the final iteration $N_3$. If only the residuals (i.e. stages 1 and 2) are used for filtering purposes, the approach is known as Residual-filtered GEP (R-GEP). If all three stages are used for filtering, the approach is known as Realizability Residual-filtered GEP (RR-GEP). 

\subsubsection{Model Training and Cost Function}

The training process was executed for 50 generations with a population size of 256, resulting in the evaluation of 6528 candidate models within the CFD-driven search space. 
The developed EARSMs were implemented only in the wake region directly behind the cylinder and the Boussinesq approximation used elsewhere, in line with best practices for developing models with improved wake flow prediction  \citep{Akolekar2019,Zhao2020}.
Once the solution reaches the final iteration $N_3$, the cost function is evaluated. The predicted solution is compared against experimentally-validated LES reference data \citep{parnaudeau2008cylinder3900}. The objective function minimizes the normalized discrepancy between two LES and RANS streamwise velocity profiles in the wake region:

\begin{equation}
J = \sum_{i=1}^2
\left[
\frac{1}{w_i}
\int_{0}^{w_i}
\left(
\frac{U^{*}_{\mathrm{LES}}(y) - U^{*}_{\mathrm{RANS}}(y)}
{\max\left(U^{*}_{\mathrm{LES}}\right)}
\right)^2
\, dy
\right],
\end{equation}
where $U^{*}$ denotes the nondimensionalized streamwise velocity, and $w_i$ represents the transverse extent of the $i$-th sampling location in the wake. The normalization by $\max(U^{*}_{\mathrm{LES}})$ ensures 
consistent weighting across sampling locations. Based on available wake data, two locations were selected for the cost function---$x/D=1.54$ and $x/D=2.02$. 

No additional penalty terms related to residual magnitude, solver divergence, or realizability are incorporated into $J$, as such constraints are strictly enforced through the preceding filtering stages. 
Models terminated during intermediate filtering stages are assigned a large penalty value 
\textit{J}, ensuring that they are not selected for propagation into the next evolutionary generation.

\section{\label{sec:ResultsDisc}Results and Discussions}

This section evaluates the performance of the filtered CFD-driven framework in terms of computational efficiency, solver convergence behavior, cost function and model coefficient formulations. The predictive capability of the learned models is demonstrated through improved wake flow  and realizability predictions. The robustness and generalization of the developed closures are further assessed using three additional test cases that are substantially different from the training cylinder flow.

\subsection{Filtered CFD-Driven Framework Performance}
The performance of the proposed filtered CFD-driven learning framework (both R-GEP and RR-GEP type) is compared with B-GEP in terms of computational speed, cost function, numerical stability (residual), model rejection rate, and physical realizability. All statistics reported in this section correspond to the evolutionary search performed for the Re = 3900 cylinder flow configuration. A fully converged cylinder flow solution, obtained using the Boussinesq approximation, is used as the starting solution from which all EARSMs are initialized.

\subsubsection{Computational Speed }
In the conventional CFD-driven B-GEP framework, each candidate closure  is allowed to proceed to full convergence typically 5,000 iterations. 
In the present filtered framework, the stage-1 check is implemented at $N_1$ =1000, stage-2 and stage-3 checks are implemented at $N_2$ = 2000, and if the models clear all three stages, the simulations are run till $N_3$ = 5000. 

\begin{figure}
\centerline
{\includegraphics[width=0.5\textwidth]{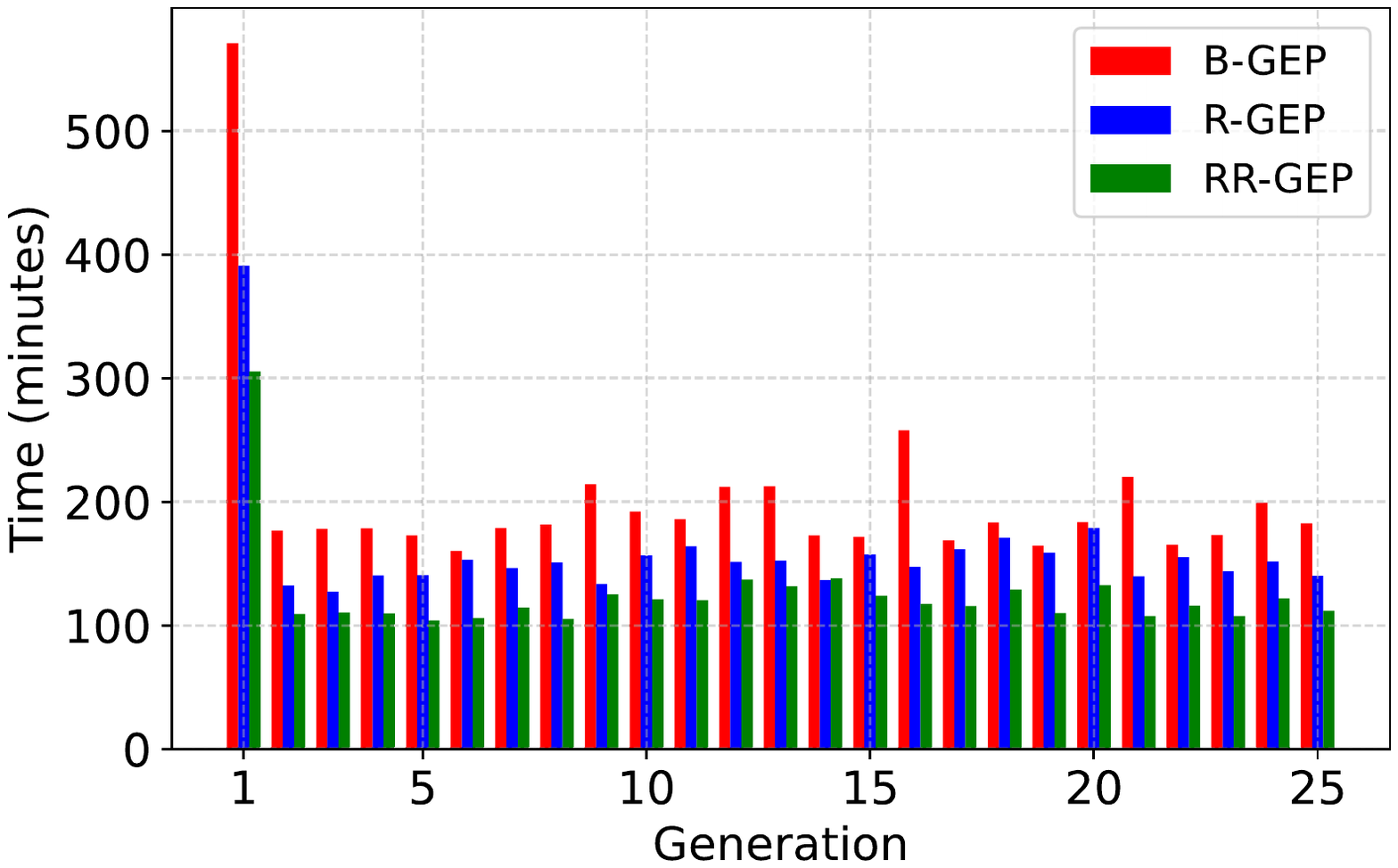}}
\caption{\label{compSpeed} Comparison of computational speed for 25 generations of B-GEP, R-GEP and RR-GEP cases.}
\end{figure}

Figure \ref{compSpeed} depicts the computation time in minutes for 25 generations for each of the B-GEP, R-GEP, and RR-GEP cases. For clarity and brevity, the figures in this section present results up to 25 generations, although all models were evolved for a total of 50 generations.  For evaluating  6528 models over 50 generations, 
B-GEP takes 23159 CPU hours, and RR-GEP takes 13371 CPU hours, representing a 42.3\% reduction in computational time in training. R-GEP reduces computational cost compared with the baseline B-GEP  by about 26.6\%. This shows that realizability filtering is the more dominant factor in reducing computational time as it prevents unstable and non-realizable models from progressing to further iterations. B-GEP exhibits the largest computational variability, with occasional spikes in runtime caused by numerically unstable (not necessarily divergent) candidate closures that run for many solver iterations before being discarded. Whereas the filtered R-GEP and RR-GEP frameworks stabilize computational cost across generations by terminating such candidates early.  The first generation requires evaluating the full population (256 individuals), whereas subsequent generations evaluate only newly generated offspring, with survivors retaining prior fitness values. With a mating probability of 0.5, this corresponds to roughly 128 evaluations per generation, reducing computational cost. In the first generation, this yields a 32\% reduction in computation time for R-GEP and 48.3\% for RR-GEP relative to B-GEP.

\begin{figure}
\centerline
{\includegraphics[width=0.5\textwidth]{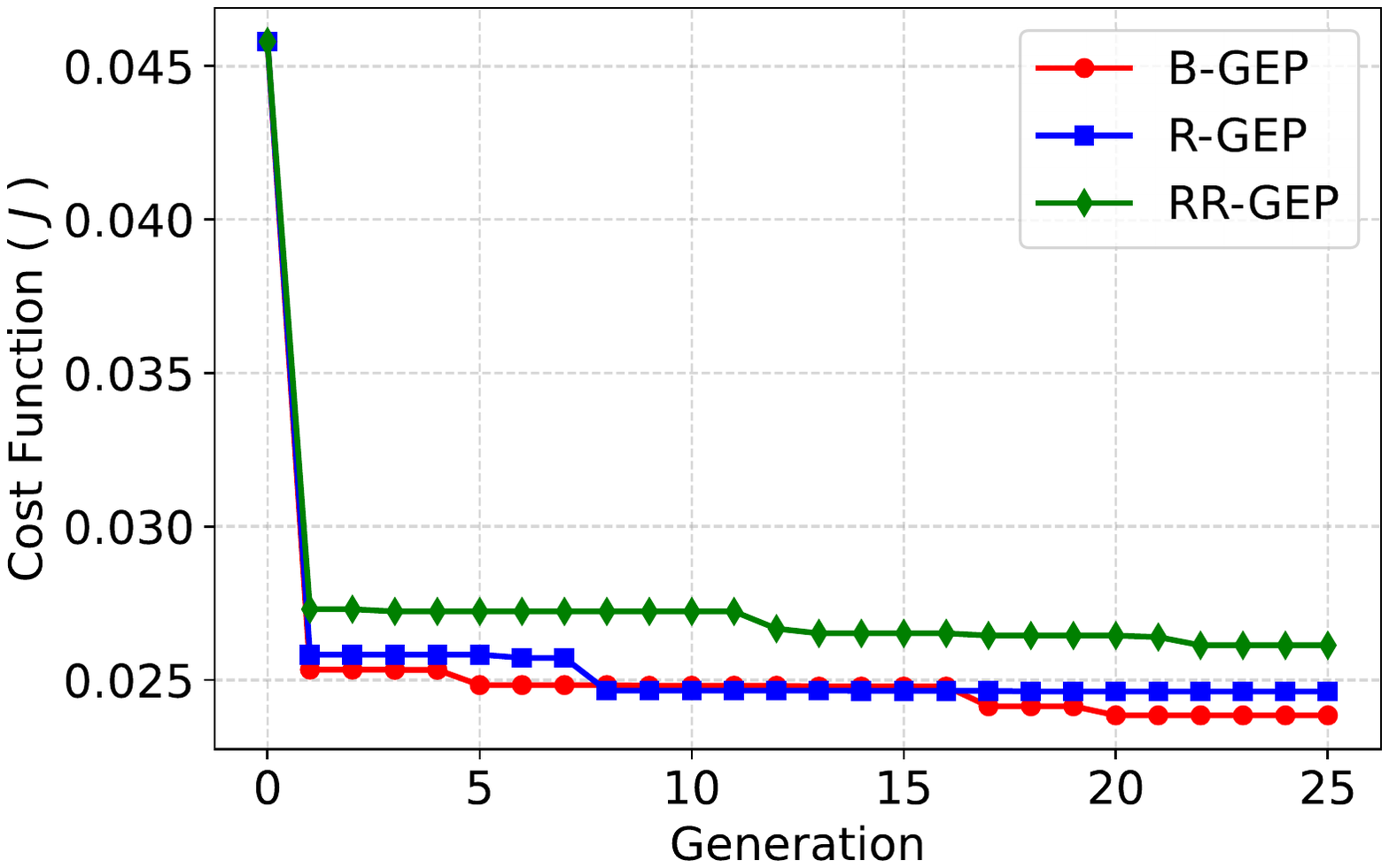}}
\caption{\label{fig:min_fit} Cost function evaluation over 25 generations.}
\end{figure}

\begin{figure*}[!t]
\centering
\includegraphics[width=1 \textwidth]{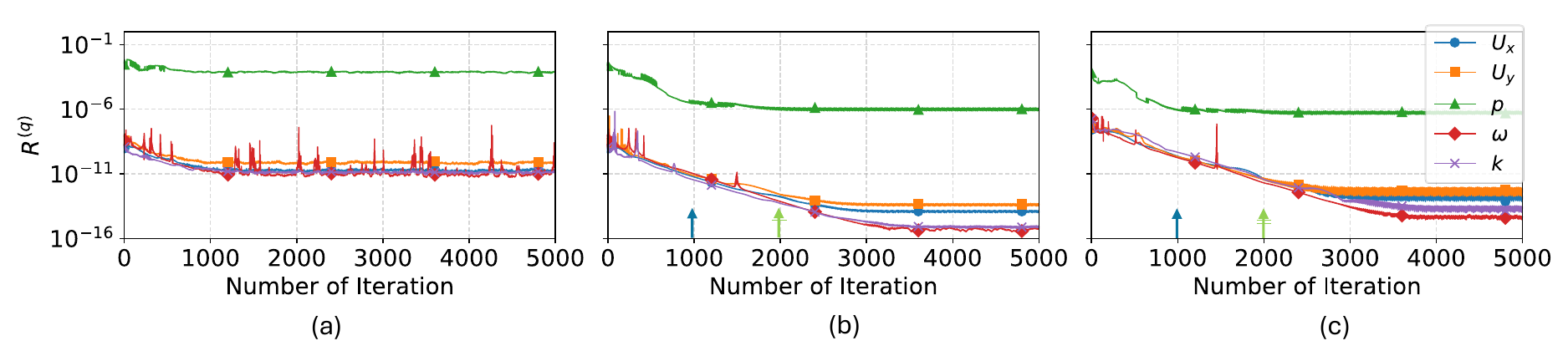}
\caption{\label{fig:residual_plot} Residual plots of the best performing model in terms of cost function (a) B-GEP, (b) R-GEP, and (c) RR-GEP.}
\end{figure*}

\begin{figure*}[t!]
\centering
\includegraphics[width=1 \textwidth]{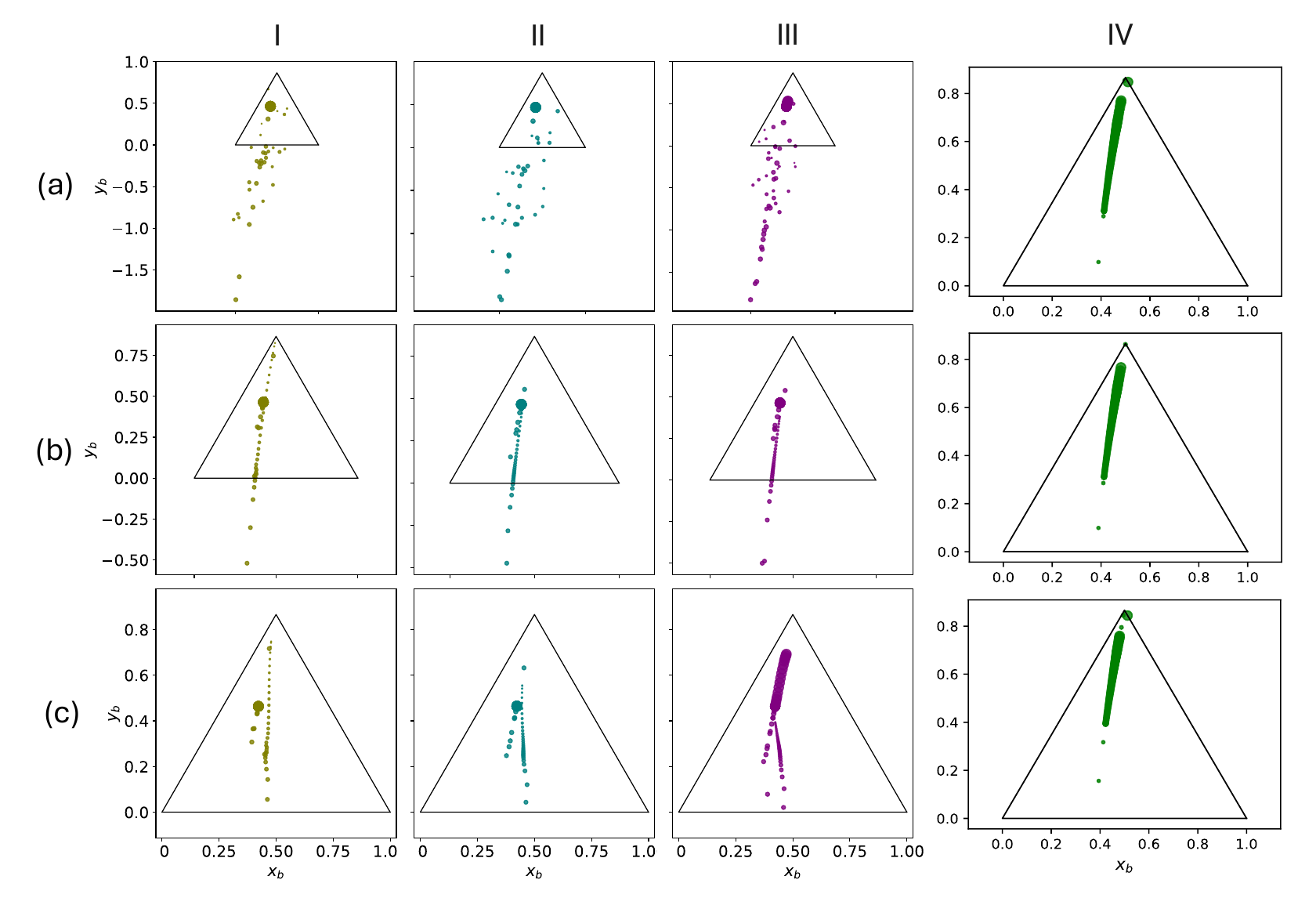}
\caption{\label{fig:barycentric} Barycentric map along three locations in the wake (I-III) and along the wake centreline (IV) for (a) B-GEP, (b) R-GEP, (c) RR-GEP.}
\end{figure*} 

\subsubsection{Cost Function}
Figure \ref{fig:min_fit} shows the evolution of the cost function over 25 generations. The cost function is defined solely in terms of wake velocity characteristics and does not explicitly account for residual convergence or realizability. The baseline RANS model with the Boussinesq approximation yields a cost function value of approximately $J_{RANS}=0.046$. A sharp reduction in cost is observed within the first generation for all methods, indicating that the evolutionary process rapidly identifies effective structural features in the closure models. This rapid improvement is also aided by the relatively large initial population size compared to earlier studies.
Beyond approximately 8–12 generations, the rate of improvement diminishes for all cases, suggesting that the search has converged to a near-optimal region of the model space, with subsequent generations providing only marginal gains. The baseline B-GEP approach achieves the lowest cost function values, reflecting its greater flexibility due to the absence of physical constraints, which allows it to exploit highly nonlinear structures to better fit the training data.

In contrast, both R-GEP and RR-GEP converge to slightly higher cost values, illustrating a trade-off between predictive accuracy and physical admissibility. The RR-GEP models stabilize at a cost of approximately 0.026, while B-GEP achieves values about 5\% lower. 
Importantly, the relatively small difference in cost function values demonstrates that these constraints do not significantly compromise predictive performance while substantially improving robustness. At 50 generations, the cost function decreases only marginally—by about 1–2\% relative to 25 generations.

\subsubsection{Numerical Stability}

Figure \ref{fig:residual_plot} shows the evolution of residuals with iteration number for the best-performing models in terms of least cost function,  obtained from (a) B-GEP, (b) R-GEP, and (c) RR-GEP. In the B-GEP case, the model with the lowest cost function exhibits relatively high pressure residuals of the order of $10^{-3}$, indicating incomplete convergence of the pressure equation. As the iterations progress, the pressure residual shows little reduction and eventually stagnates, suggesting persistent pressure–velocity imbalance. This stagnation is also reflected in numerical oscillations in the pressure field after approximately 5000 iterations, despite the other variables converging to values close to 
$10^{-10}$.

In contrast, both R-GEP and RR-GEP demonstrate significantly improved convergence behavior. The pressure residual decreases to approximately 
$10^{-6}$, indicating a much more stable solution, while the remaining variables converge below $10^{-12}$. The RR-GEP model achieves slightly lower pressure residual levels than R-GEP, suggesting that the additional filtering stage further enhances numerical stability. These results highlight the effectiveness of residual-based filtering in removing candidate models that lead to poor solver convergence.
It was also noted that realizability violations were frequently preceded by elevated pressure residuals, indicating strong coupling between anisotropy instability and pressure–velocity imbalance. 

\subsubsection{Realizability}

Figure \ref{fig:barycentric} shows the barycentric maps along three locations in the wake (I--III) and along the wake centreline (IV), as illustrated in Figure \ref{fig:cylinder_schematic}, for (a) B-GEP, (b) R-GEP, and (c) RR-GEP. The results clearly demonstrate a progressive improvement in realizability from B-GEP to RR-GEP. For locations I--III, the marker size increases with distance away from the wake centreline. 
The B-GEP models exhibit significant scatter and frequent excursions outside the realizable domain, indicating non-physical anisotropy states. In contrast, the introduction of residual-based filtering in R-GEP substantially reduces these violations. For example, the maximum violation reduces from $y_B = -2.0$ in B-GEP to approximately $y_B = -0.5$ in R-GEP. Although this represents a significant improvement, some dispersion and localized realizability violations persist.
The RR-GEP models, however, produce tightly clustered and well-aligned distributions across all wake locations, indicating bounded and consistent anisotropy evolution. Quantitatively, the variance perpendicular to the principal anisotropy trajectory is reduced by approximately an order of magnitude in RR-GEP compared to B-GEP.
Perpendicular to the wake centreline, the turbulence exhibits near-isotropic behavior close to the wake centreline, becomes increasingly anisotropic within the shear layers, and gradually returns toward isotropy in the free-stream region. 

Along the wake centreline (IV), all models remain within the realizable domain due to the symmetric and weakly sheared nature of the flow, which promotes bounded and near-isotropic turbulence states even for less stable closures. Moving downstream from the trailing edge (increase in marker size from cylinder), the turbulence progressively approaches a more isotropic state. This demonstrates the effectiveness of the combined residual and realizability filtering in enforcing physically admissible and numerically stable turbulence closures.

\subsubsection{Model Rejection Behavior}

\begin{figure}
\centerline
{\includegraphics[width=0.5\textwidth]{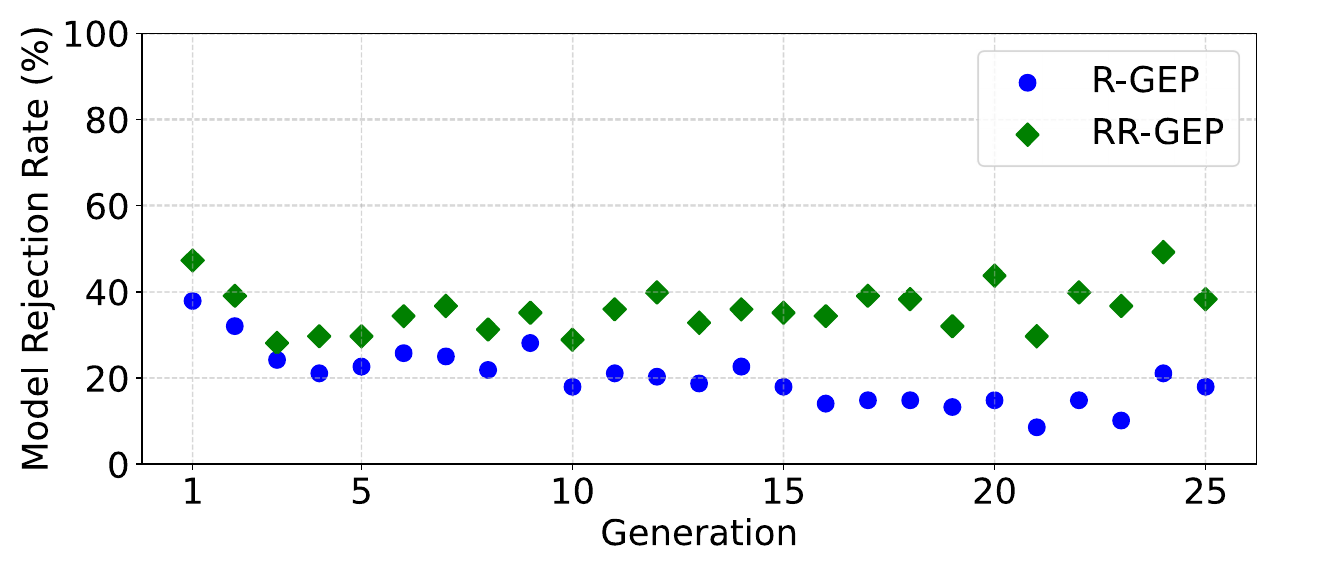}}

\caption{\label{fig:kill} Model rejection rate for R-GEP and RR-GEP with generation. }
\end{figure}

Figure \ref{fig:kill} shows the model rejection rate for R-GEP (stages 1,2)  and RR-GEP (for stages 1,2,3) with respect to the number of generations. In the first generation, R-GEP and RR-GEP reject 37.9\% and 47.26\% of the models. Across 50 generations, the average rejection rate for R-GEP is 21.5\% and for RR-GEP is 37.8\%. The model rejection rate trends thus highlight the role of filtering in stabilizing the evolutionary search. RR-GEP consistently rejects a larger fraction of candidate models than R-GEP, indicating that the additional realizability-based screening effectively identifies non-realizable closures early. 

\begin{figure}
\centerline
{\includegraphics[width=0.5\textwidth]{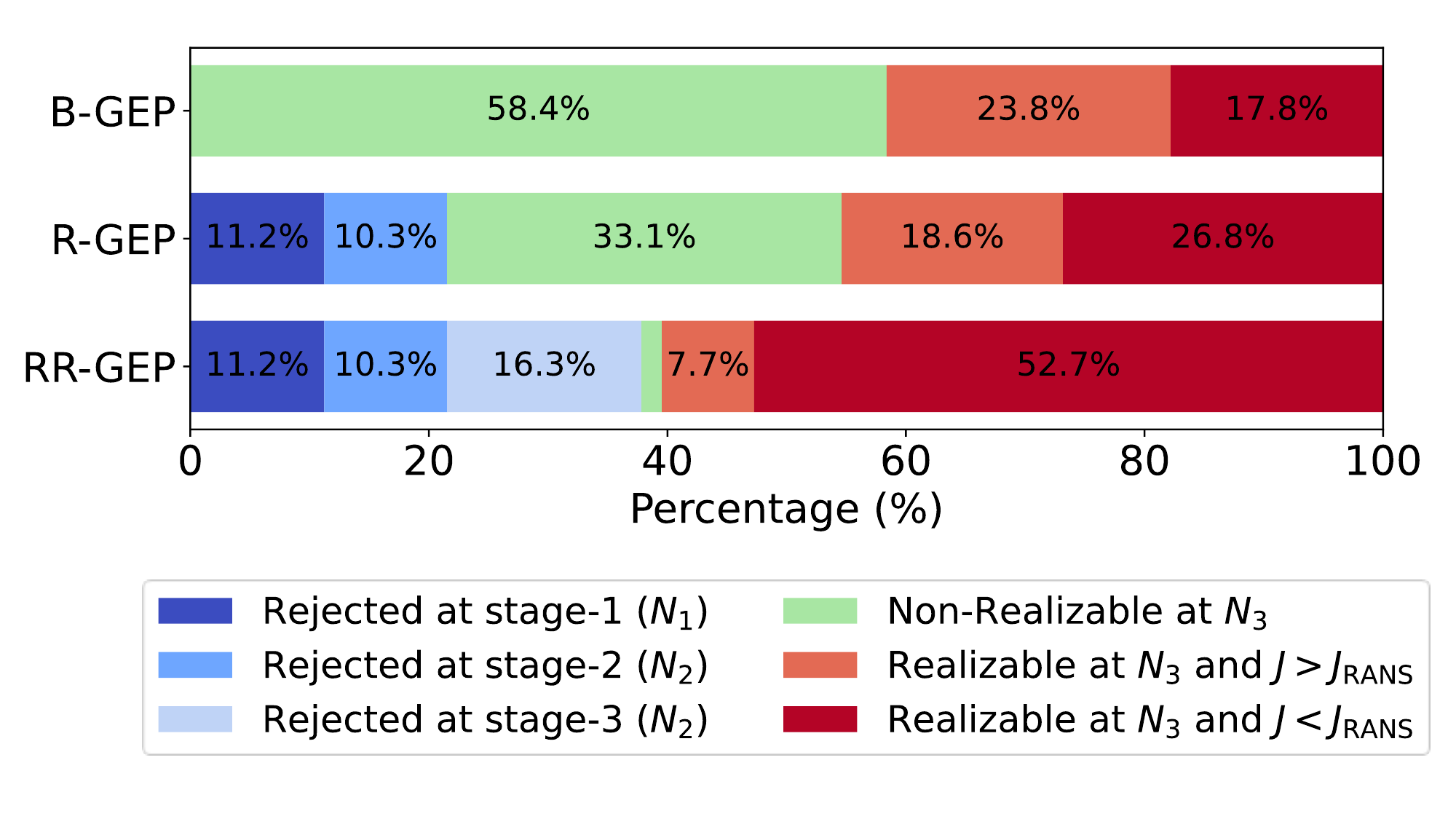}}
\caption{\label{fig:rej_per} Model class distribution and rejection behavior across B-GEP, R-GEP, and RR-GEP cases.}
\end{figure}

Figure~\ref{fig:rej_per} presents the distribution of model classes and rejection behavior for B-GEP, R-GEP, and RR-GEP across 50 generations, expressed as a percentage of the total candidate models. For B-GEP, at \(N_3 = 5000\) iterations, a majority of the models (58.4\%) remain non-realizable. Among the realizable candidates, only 17.8\% achieve a cost function lower than the baseline RANS value (0.046), while none reduce it below 40\% of the baseline RANS (0.0276). The remaining 23.8\% of models, although realizable, perform worse than the RANS baseline.

Stage--1 (\(N_1\)) eliminates approximately 11.2\% of models, followed by an additional 10.3\% rejection at Stage--2 (\(N_2\)). In RR-GEP, a further 16.3\% of candidates are rejected in Stage--3 (\(N_2\)). At \(N_3\), R-GEP still retains a considerable fraction of non-realizable models (33.1\%). Among the realizable set, 26.8\% outperform the RANS baseline, while the remainder exhibit higher cost function values. In contrast, RR-GEP yields a near-complete elimination of non-realizable models, with only 1.7\% remaining at \(N_3\). These correspond to temporary realizability violations tolerated at \(N_2\). Notably, the majority of RR-GEP models achieve lower cost function values than the baseline RANS. Overall, these results indicate that enforcing realizability constraints at intermediate stages---particularly at \(N_2\)---is highly effective: models satisfying these constraints are overwhelmingly likely to remain realizable upon full convergence at \(N_3\), while also delivering improved predictive performance.

\begin{figure}
\centerline
{\includegraphics[width=0.5\textwidth]{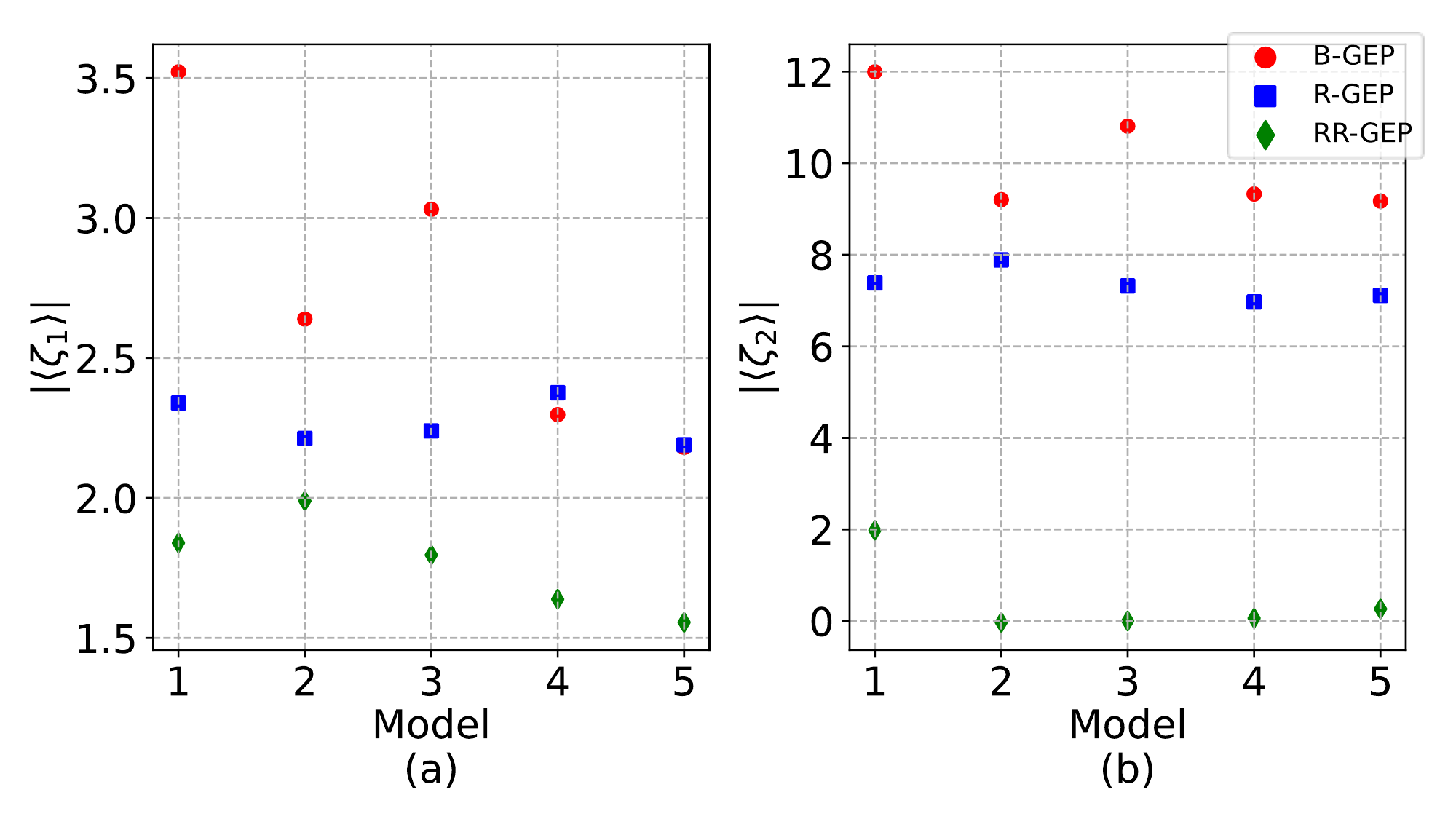}}
\captionof{figure}{ Absolute value of mean (a) $\zeta_1$ and (b) $\zeta_2$ coefficients for five best performing models at $x/D = 2.02$ for y=0 to y/D = 0.91.}
\label{fig:zetas}
\end{figure}

\subsubsection{{Model Formulation}} 

The differences between the learned closures are examined by analyzing the coefficient distributions of the tensor basis functions and the resulting anisotropy components. Figure~\ref{fig:zetas} shows the magnitude of the mean values of $| \langle \zeta_1 \rangle |$ ($\zeta_1$ is negative) and $| \langle \zeta_2 \rangle | $ ( $\zeta_2$ is positive) of five best-performing models, at $x/D = 2.02$ averaged from $y/D=0$ to $y/D=0.91$, for the B-GEP, R-GEP, and RR-GEP formulations. 
The coefficient trends reveal a clear progression in model behavior. The B-GEP models exhibit relatively large coefficient magnitudes, with an average value of $ | \langle \zeta_1\rangle|\approx 2.73$ and $|\langle \zeta_2\rangle |\approx 10.1$ across the five models. This indicates that strong nonlinear contributions from the tensor basis. The R-GEP models show moderated coefficients, with an average value of  $ |\langle \zeta_1 \rangle |  \approx 2.27$ and $ |\langle \zeta_2\rangle|  \approx 7.33$, reflecting the influence of residual-based filtering in suppressing unstable growth. In contrast, the RR-GEP models yield significantly reduced and tightly bounded coefficients, with an average value of $ |\langle \zeta_1\rangle|  \approx 1.76$ and $ |\langle \zeta_2\rangle|  \approx 0.45$, indicating a preference for simpler and more controlled functional forms. These differences directly translate into the predicted anisotropy structure.

\begin{figure}
\centerline
{\includegraphics[width=0.5\textwidth]{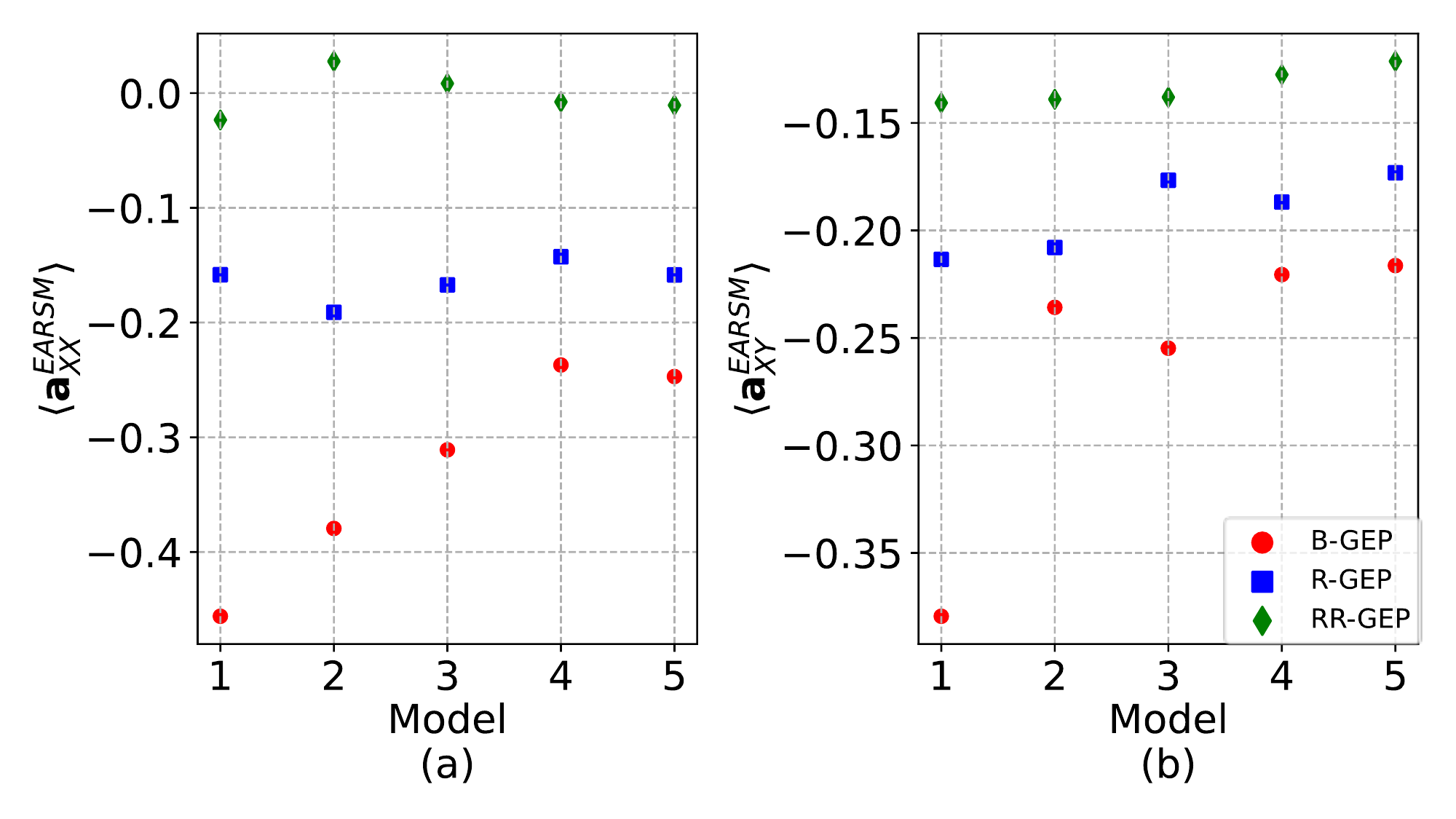}}
\caption{\label{fig:aij} Mean (a) $\bm a^{EARSM}_{XX}$ and (b) $\bm a^{EARSM}_{XY}$ for five best performing models at $x/D = 2.02$ for y/D=0 to y/D = 0.91.}
\end{figure}

Figure~\ref{fig:aij} presents the corresponding mean anisotropy components $\bm a^{\text{EARSM}}_{XX}$ and $\bm a^{\text{EARSM}}_{XY}$ at the same location. The B-GEP models produce elevated normal anisotropy levels, with $ \langle \bm a^{\text{EARSM}}_{XX} \rangle \approx -0.32$, indicating excessive streamwise stress. The R-GEP models reduce this by yielding intermediate values of $\langle \bm a^{\text{EARSM}}_{XX}\rangle  \approx -0.16$. In contrast, the RR-GEP models predict near-zero values, $ \langle \bm a^{\text{EARSM}}_{xx} \rangle \approx -0.001$, consistent with physically realizable wake turbulence where anisotropy is primarily shear-driven rather than dominated by streamwise normal stresses. The shear component, $ \langle \bm a^{\text{EARSM}}_{XY}\rangle$, follows a similar trend and is the dominant contributor across all models. The RR-GEP formulation maintains lower shear stress levels leading to a more balanced representation of turbulence production and redistribution. The B-GEP formulation, due to its stronger nonlinear coupling between invariants ($I_1$, $I_2$) and tensor basis functions, amplifies both normal and shear stresses, pushing the anisotropy tensor toward unrealizable states. 
Such trends in the model coefficients and anisotropy components are also visible at all other locations. There is thus a clear trend in the model coefficients and anisotropy components which can directly be linked to the residual and realizability-based filtering.

The model coefficients with the least cost function (after 50 generations) from each of the three formulations are shown here: 

\noindent B-GEP:
\begin{equation}
\begin{aligned}
\zeta_1 &= -2 I_1 + I_2 - 3.36, \\
\zeta_2 &= 3 I_1 - I_2 + 11.77, \\
\zeta_3 &= 0.82 + 0.37 I_1 + 0.37 I_1^2 + 1.63 I_1 I_2 + 2 I_1^2 I_2 - 2 I_1 I_2^2, \\
\zeta_4 &= 2 I_1 + 0.1 I_2.
\end{aligned}
\end{equation}

\noindent R-GEP:
\begin{equation}
\begin{aligned}
\zeta_1 &= -2.34 - I_1 I_2, \\
\zeta_2 &= 7.39 - I_1 - I_2, \\
\zeta_3 &= 3.34 - I_1 + I_2, \\
\zeta_4 &= 0.73 I_1 - 0.73 I_2.
\end{aligned}
\end{equation}

RR-GEP:
\begin{equation}
\begin{aligned}
\zeta_1 &= -1.84 - 0.02 I_1, \\
\zeta_2 &= 2 + I_2 - I_1^2, \\
\zeta_3 &= 2.41 - 2.22 I_1 - 1.13 I_2 \\ 
\zeta_4 &= I_1 - 3.15.
\end{aligned}
\end{equation}

\subsection{Cylinder Flow Characteristics}
\begin{figure}

{\includegraphics[width=0.5\textwidth]{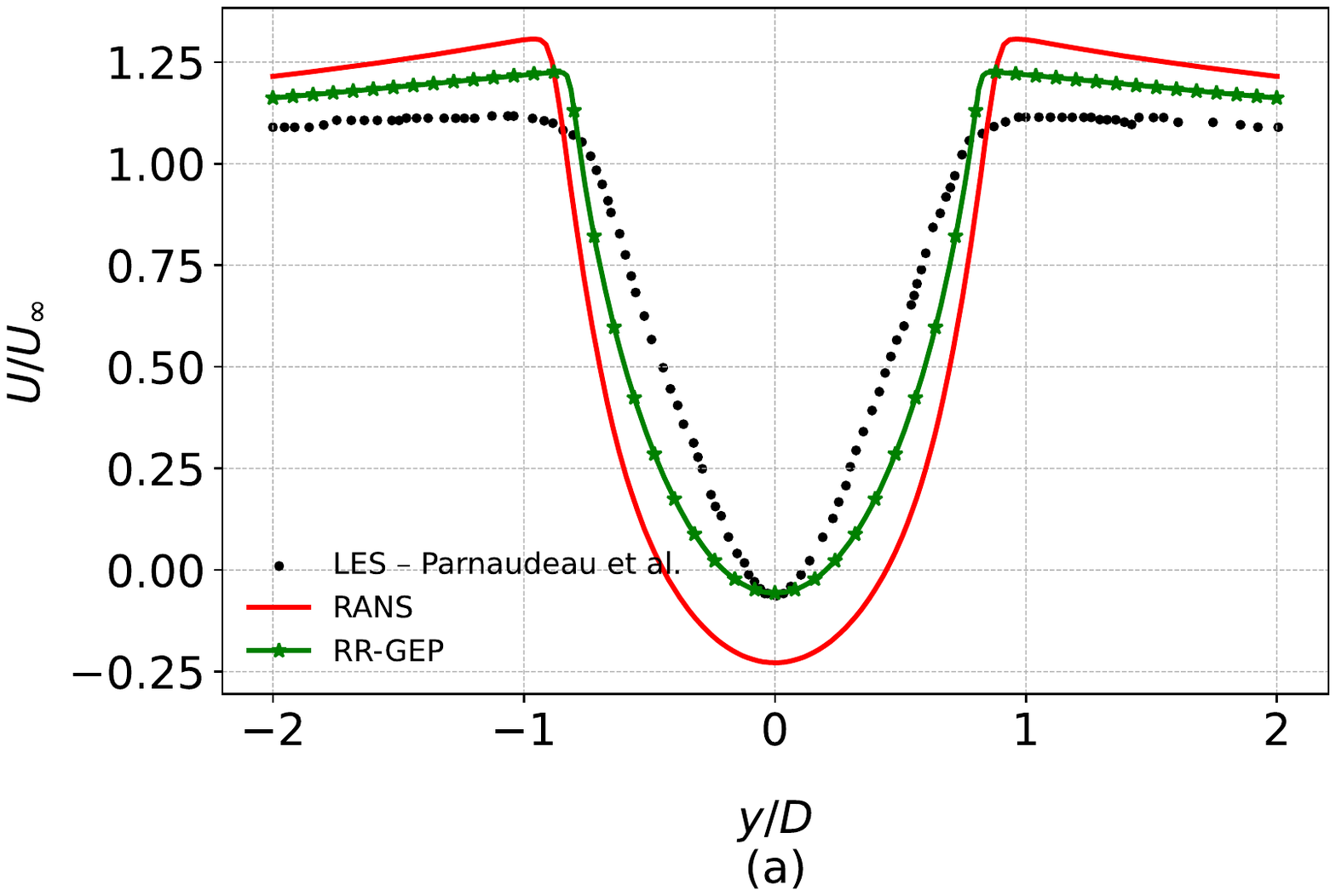}}
{\includegraphics[width=0.5\textwidth]{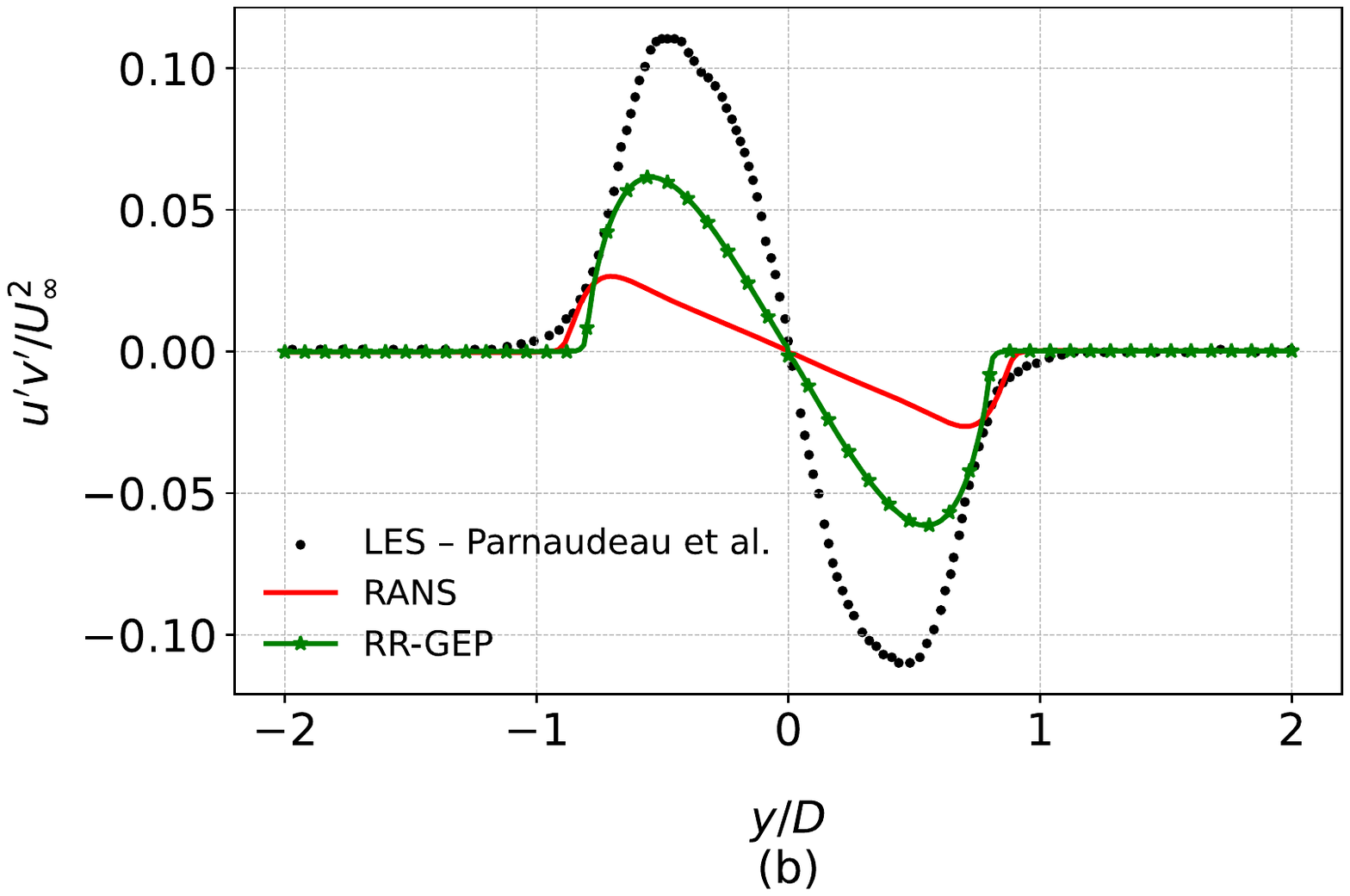}}
{\includegraphics[width=0.5\textwidth]{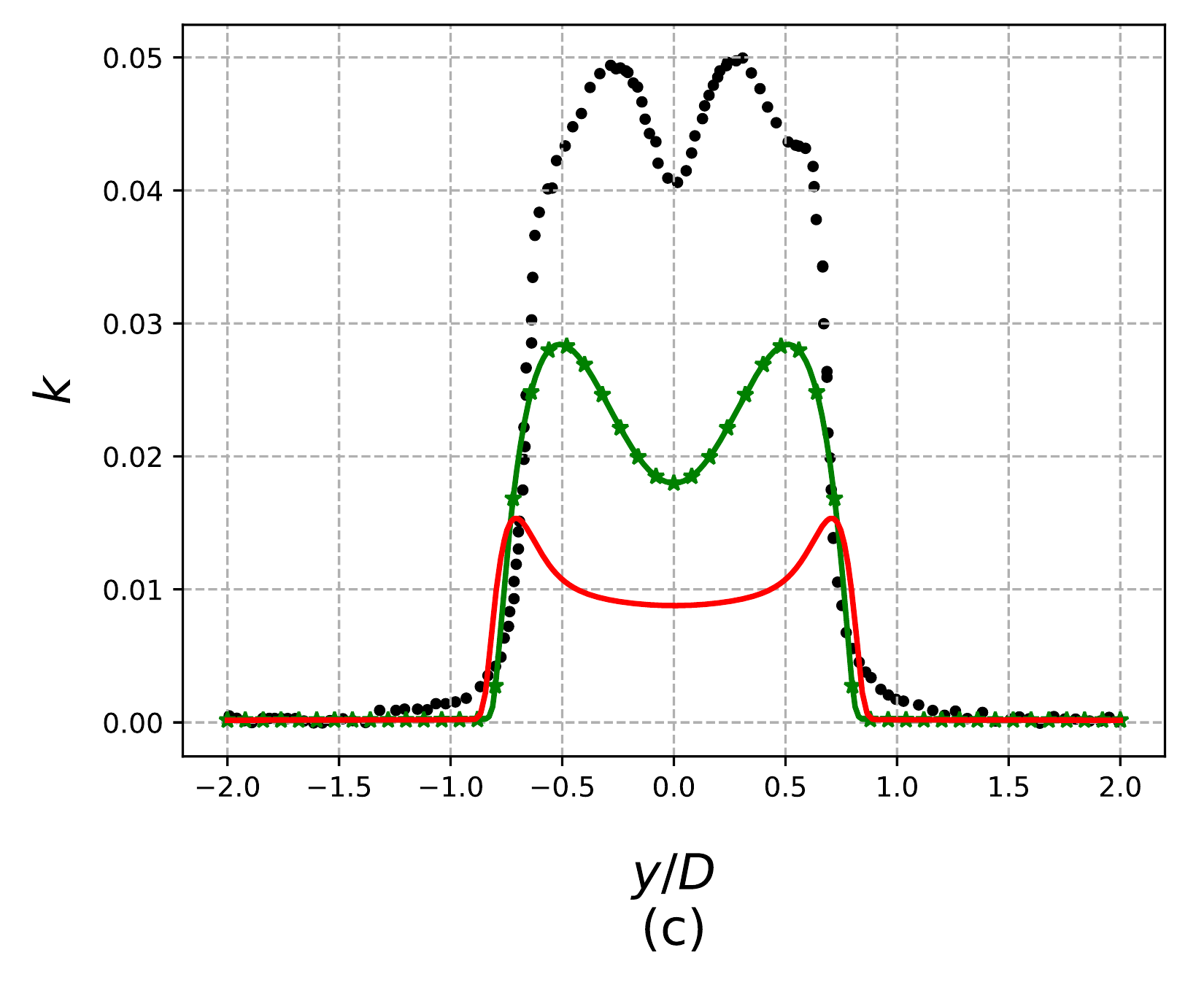}}
\caption{\label{fig:u_xd202}  (a) Streamwise velocity profile ($U/U_{\infty}$) and (b) Reynolds shear stress ($u'v'$) at $x/D= 2.02$.}
\end{figure}

This section presents the performance of the best RR-GEP model in capturing flow characteristics, compared against baseline RANS ($k\omega$–SST with the Boussinesq approximation) and LES \citep{parnaudeau2008cylinder3900}. 
Figure \ref{fig:u_xd202} (a) shows the streamwise velocity profile at $x/D = 2.02$.
The baseline RANS model overpredicts the wake width and exhibits a larger velocity deficit relative to LES. In contrast, the RR-GEP model yields a narrower wake and reduced deficit, achieved through enhanced turbulent diffusion and a smaller separation bubble. Consequently, the RR-GEP model provides a more physically consistent solution and improved agreement with LES across the wake region.
The baseline RANS model exhibits excessive turbulent diffusion, resulting in an overly dissipated wake characterized by a narrow velocity deficit and accelerated recovery. In contrast, the RR-GEP model significantly improves the wake structure, yielding fuller velocity profiles and better agreement with reference data at both sampling locations. The improvement is particularly evident in the near-wake region, where shear-layer development and momentum redistribution are more accurately captured.

Although Reynolds stress components are not explicitly included in the cost function, the RR-GEP model demonstrates notable improvement in both normal and shear stress predictions. The $u'v'$ profiles, as shown in Fig. \ref{fig:u_xd202}(b) show a more accurate peak magnitude and spatial distribution, indicating improved representation of turbulence production and transport mechanisms. The shear stress component is the most important component in free-shear flows and the peak non-dimensionalized shear stress magnitude increases from 0.025 in the baseline RANS case to 0.055 in the RR-GEP case which is much closer to the peak LES value of 0.11.  There is also a significant improvement in the TKE over the baseline RANS model.



\begin{figure}
\centerline
{\includegraphics[width=0.6\textwidth]{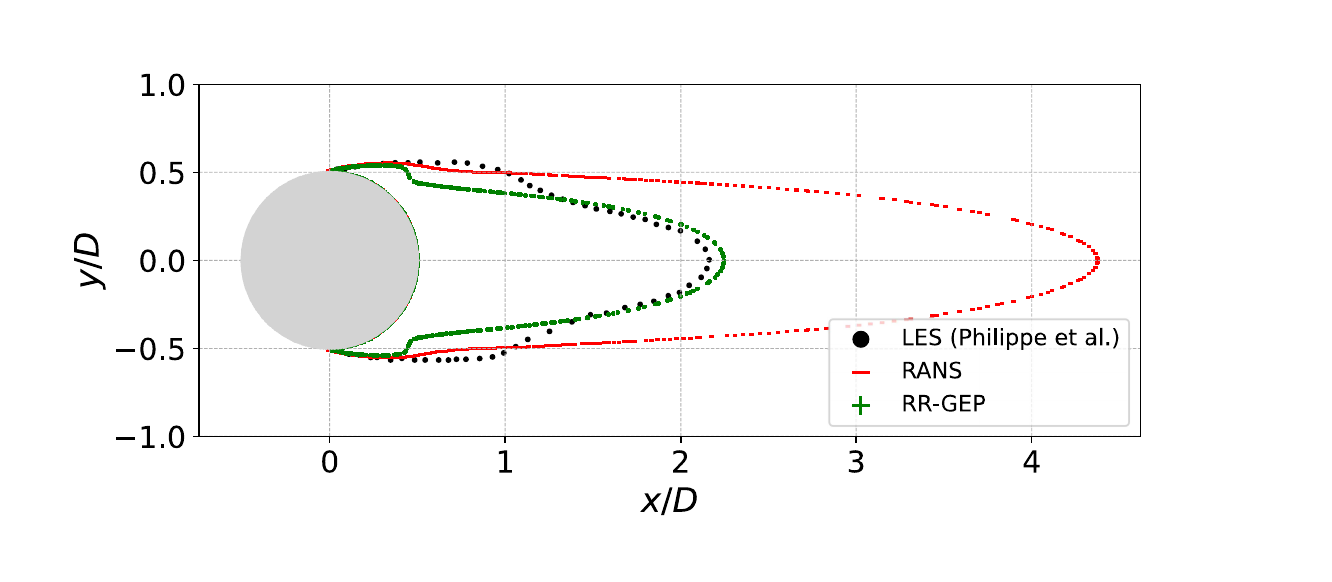}}
\caption{\label{fig:sepbubble} The separation bubble in the wake region of the cylinder at Re=3900.}
\end{figure}

The enhancement in closure prediction directly translates to improved prediction of the separation bubble. From Fig. \ref{fig:sepbubble} The baseline RANS model significantly overpredicts the extent of the recirculation region. 
In contrast, the RR-GEP model yields a substantially reduced separation bubble length, closely matching the reference solution in both extent and topology. The recirculation region is more compact and physically realistic as compared to the LES case. The extent of the separation region reduces to $x/D=2.25$ in the RR-GEP case, as compared to that baseline RANS case which extends till about $x/D=4.5$. 

\subsection{Test Cases}
The RR-GEP model is further evaluated on different geometries, at Reynolds numbers and spatial locations beyond its training conditions. It consistently improves wake predictions for both bluff and streamlined bodies, while maintaining realizability throughout the wake region where the model is applied. Only one wake location is shown for brevity. 


\begin{figure}

{\includegraphics[width=0.5\textwidth]{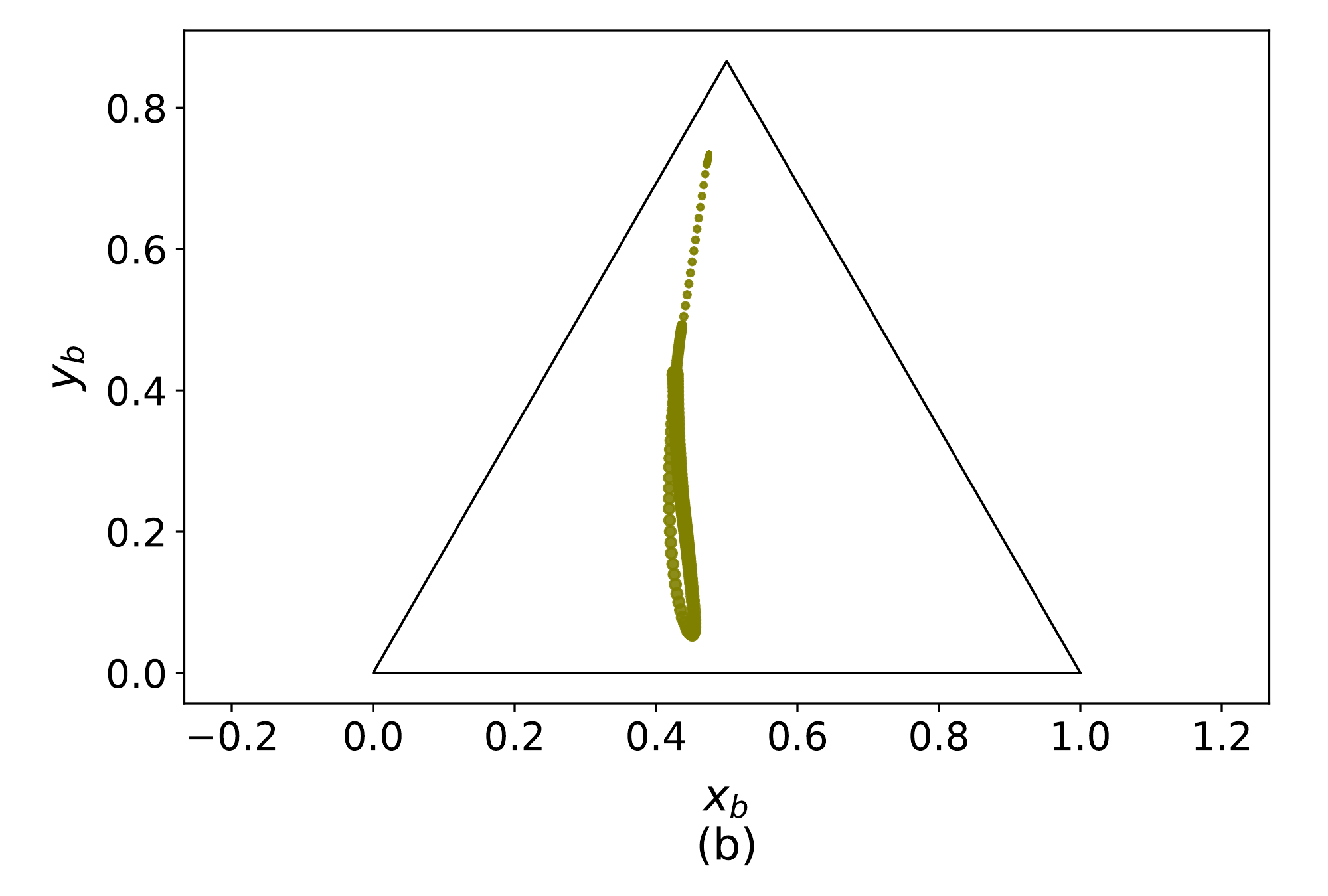}}
{\includegraphics[width=0.5\textwidth]{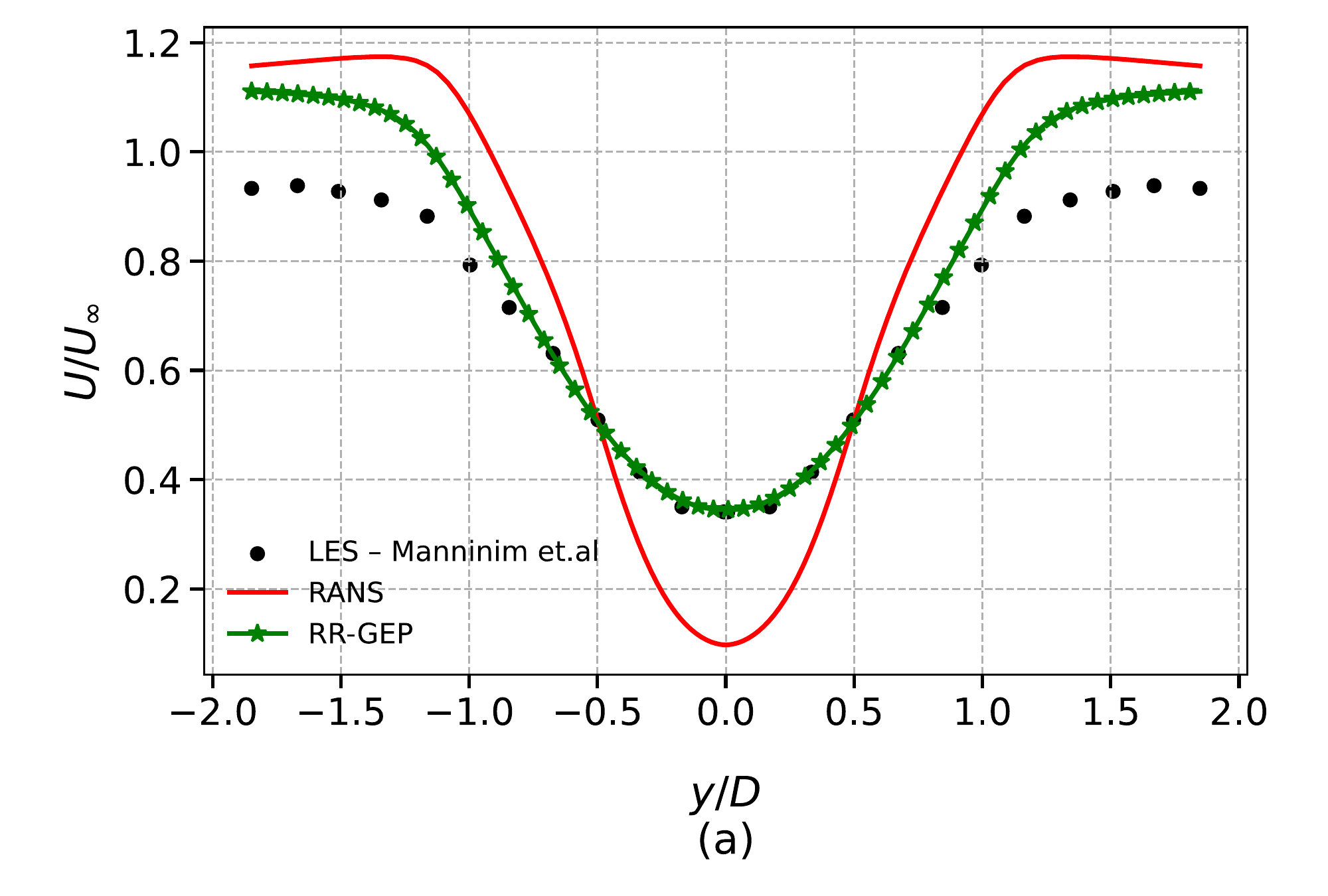}}
\caption{\label{fig:baryRect} (a) Wake profiles and (b) barycentric map for the rectangular 5:1 cylinder at $x/H=4.5$.}
\end{figure}

\subsubsection{Rectangular 5:1 Cylinder}

A Reynolds numbers of Re=44,900 is considered based on the experimental data of Mannini et al. \citep{mannini2019benchmarkRectangularCylinder}. Wake profiles at \(x/H = 4.5\) (refer Fig. \ref{fig:rect_schematic}) are shown in Figure \ref{fig:baryRect}(a). 
The baseline RANS model significantly overpredicts turbulent diffusion, leading to an excessively deep and narrow wake with a minimum normalized velocity of approximately 0.1. In contrast, the RR-GEP model increases turbulent diffusion, resulting in a fuller and wider wake profile. The predicted minimum velocity increases to approximately 0.37, closely matching the LES data for Re = 44,900. A similar level of agreement is observed for the entire wake region, including the near-wake region, where the separation bubble is more accurately captured compared to the baseline RANS model. Importantly, since the model is applied only in the wake region, the flow over the rectangular cylinder surface remains unaffected
The wake prediction also improves at higher Reynolds number case of Re=112,600.  
The corresponding barycentric map at \(x/H = 4.5\) (Fig.~\ref{fig:baryRect} (b)) from the wake-centreline to towards the freestream region, further confirms the physical consistency of the model. The RR-GEP predictions remain entirely within the realizable domain, with tightly clustered anisotropy states aligned along a physically consistent trajectory. This demonstrates that the model not only improves wake predictions but also maintains realizability. Such an observation is made for the entire wake region in case of the 5:1 cylinder. 


\subsubsection{NACA 0012 Airfoil}

\begin{figure}

{\includegraphics[width=0.5\textwidth]{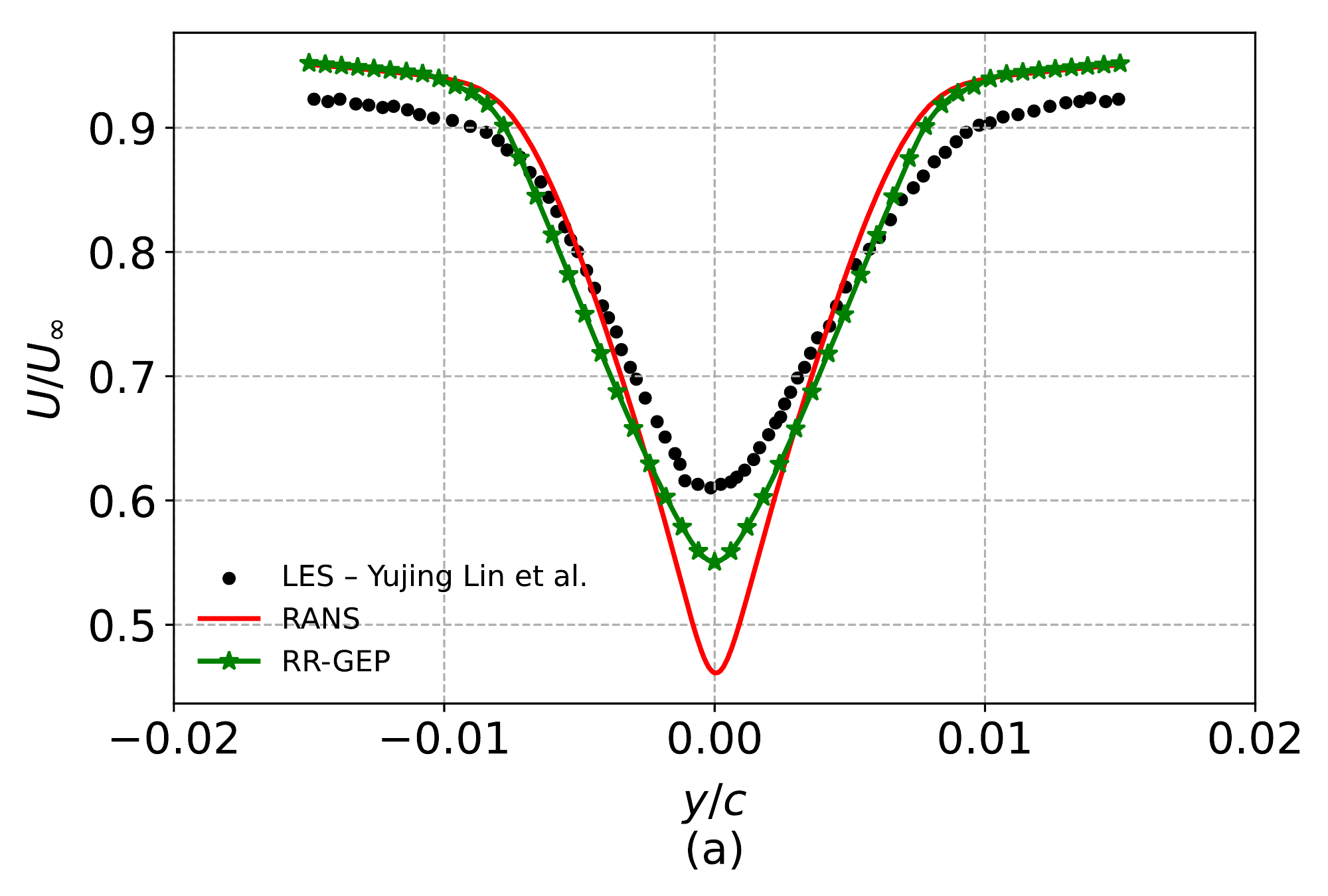}}
{\includegraphics[width=0.5\textwidth]{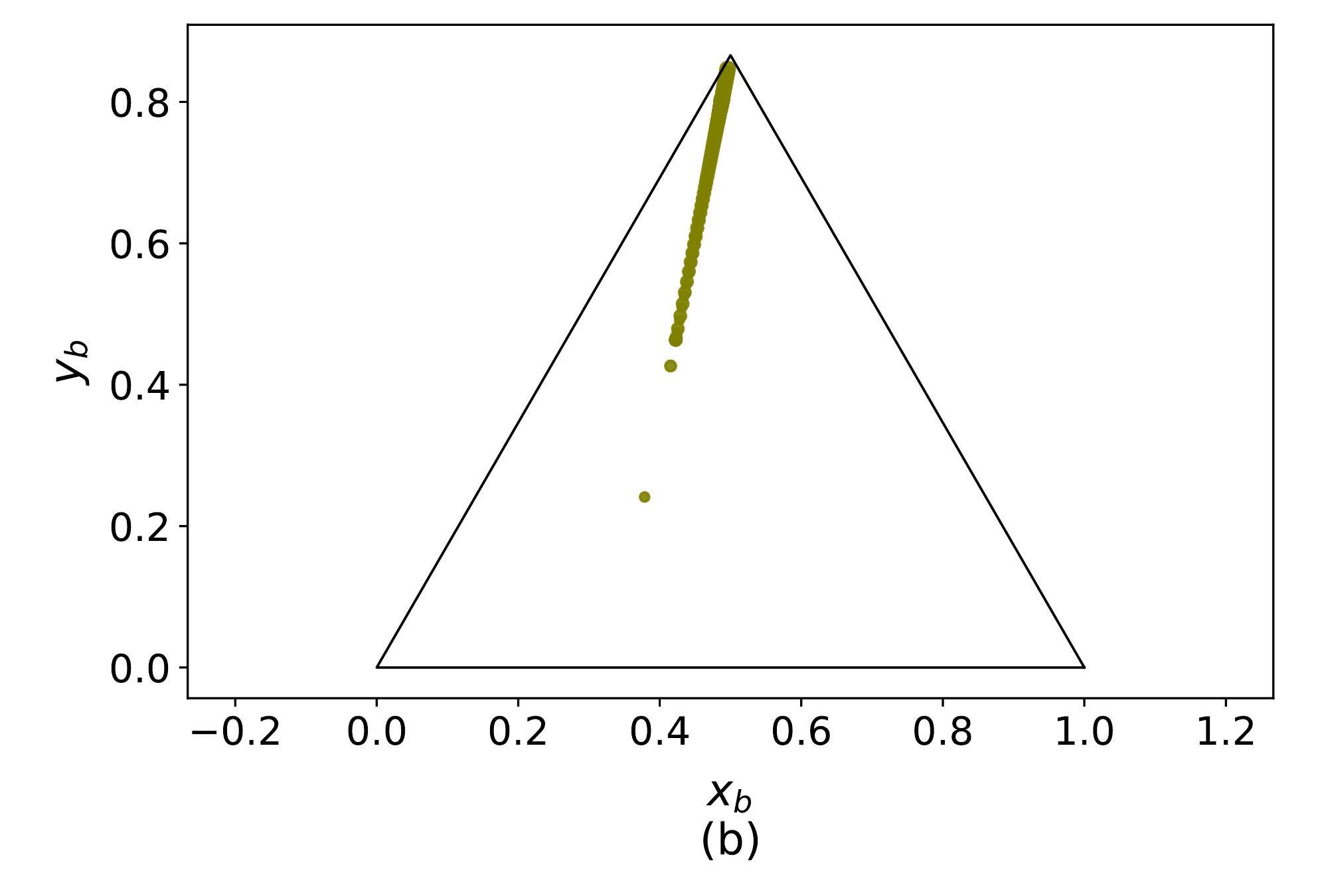}}
\caption{\label{fig:airfoilwake} (a) Wake profiles and (b) barycentric map for a  NACA 0012 airfoil at $x/C = 1.05$.}
\end{figure}

The predictive capability of the trained RR-GEP model is further assessed on a canonical streamlined configuration, namely the NACA 0012 airfoil at $\mathrm{Re}=2\times10^{5}$. This case contrasts bluff-body wakes, featuring largely attached boundary layers with mild separation, and acts a good test case to assess generalizability. Additionally, the Reynolds number is two orders of magnitude higher than the training case. Wake velocity profiles at $x/c = 1.05$ (refer to Fig. \ref{fig:naca_geometry}) are presented in Fig.~\ref{fig:airfoilwake}(a). The baseline RANS model exhibits excessive turbulent diffusion, leading to a smeared wake and accelerated velocity recovery. In contrast, the RR-GEP model preserves the wake deficit more accurately, yielding improved agreement with reference data and a more realistic representation of momentum transport in the near wake. The corresponding barycentric map at $x/c = 1.05$ (Fig.~\ref{fig:airfoilwake}(b)) highlights the improved anisotropy prediction. The RR-GEP model remains strictly within the realizable domain. 
These results demonstrate that the RR-GEP model, despite being trained on a bluff-body cylinder flow, retains robustness when applied to streamlined configurations, improving wake prediction without degrading attached-flow behavior.

\subsubsection{DARPA Suboff}
The robustness of the RR-GEP model is further evaluated on the DARPA Suboff geometry at $\mathrm{Re}=1.1\times10^{6}$, testing the performance of the model at high-Reynolds-number conditions with mild separation and complex wake recovery.
\begin{figure}
{\includegraphics[width=0.4\textwidth]{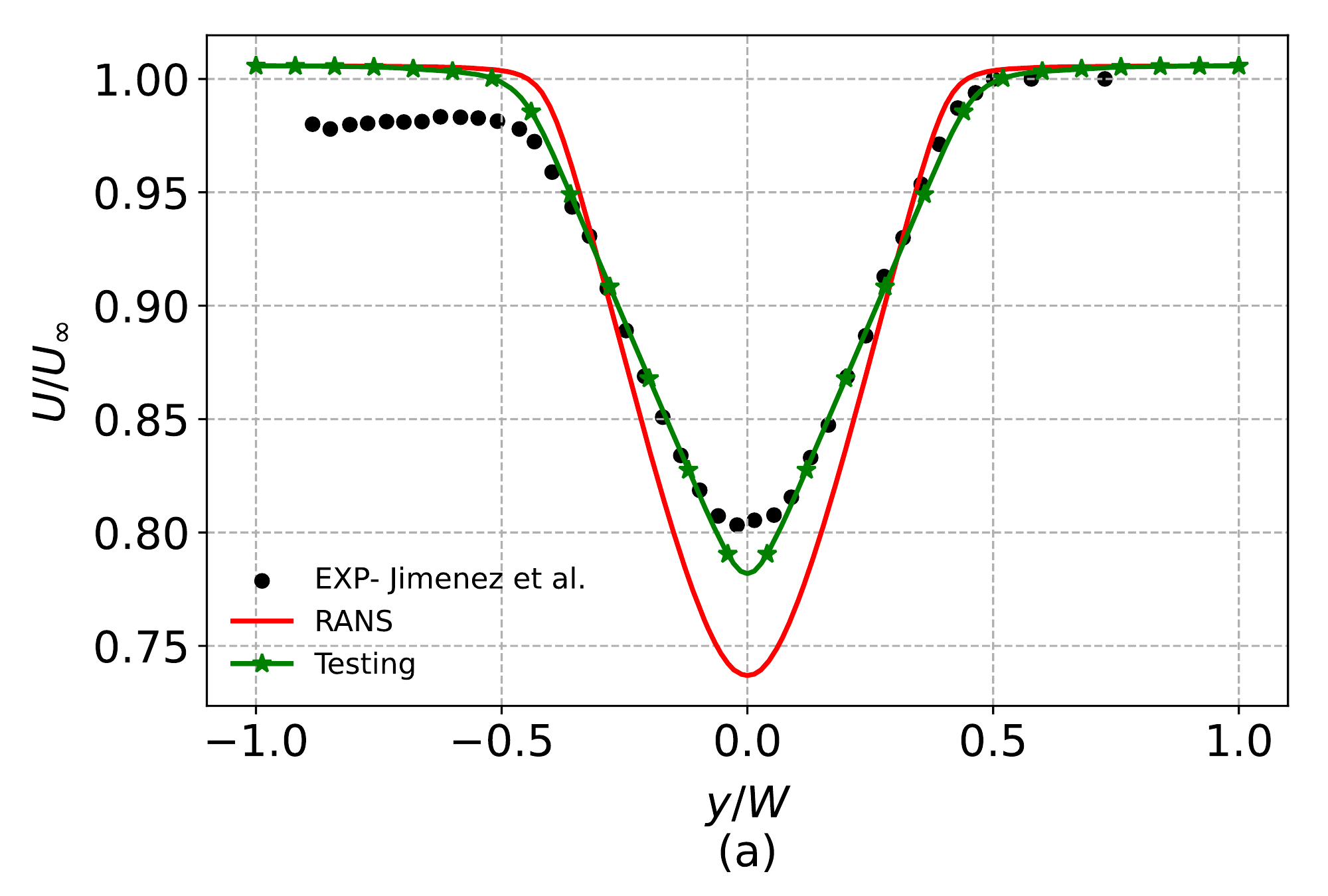}}
{\includegraphics[width=0.4\textwidth]{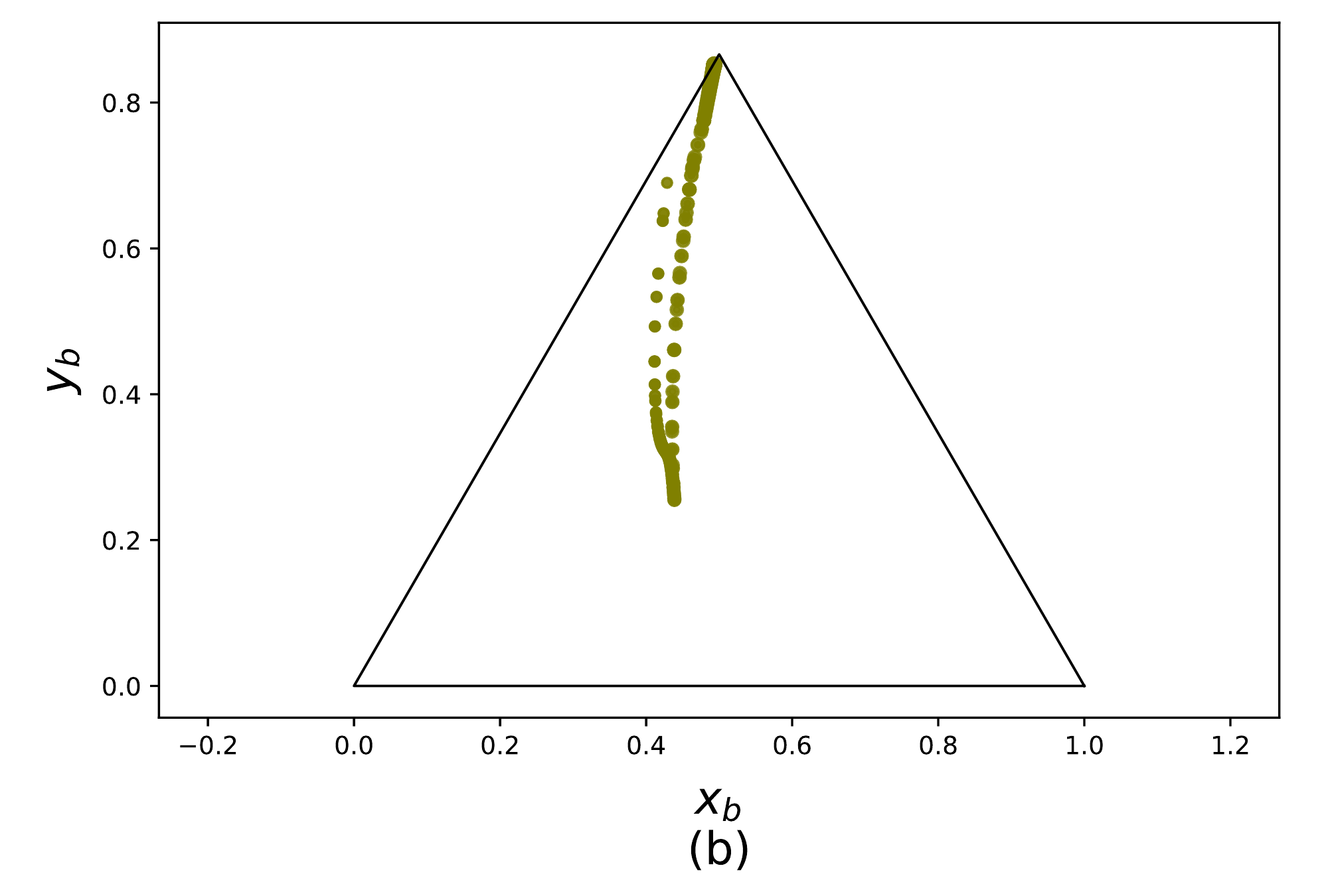}}
\caption{\label{fig:darpa_wake_xd12} (a) Wake profiles and (b) barycentric map for the DARPA Suboff at $x/W$ = 12.}
\end{figure}

Wake profiles are shown in Fig. \ref{fig:darpa_wake_xd12}(a) at $x/W=12$ (refer Fig. \ref{fig:darpa_schematic}). The baseline RANS model exhibits excessive diffusion, leading to an overprediction of the velocity deficit and overly rapid wake recovery. The RR-GEP model improves the wake prediction.
The realizability of the RR-GEP model is assessed using the barycentric map at $x/W = 12$, as shown in Figure \ref{fig:darpa_wake_xd12}(b). The predicted anisotropy states remain strictly confined within the barycentric triangle, demonstrating that the realizability constraints enforced during training are preserved even in this axisymmetric, high-Reynolds-number flow. 
These results demonstrate that the RR-GEP model, trained on a two-dimensional bluff-body flow, generalizes effectively to an axisymmetric configuration, maintaining both predictive accuracy and physical realizability across fundamentally different flow regimes and Reynolds number which is three orders of magnitude higher. 

\section{Conclusions}

A residual- and realizability-filtered CFD-driven machine learning framework has been developed to address key limitations in turbulence model discovery using symbolic regression. The proposed approach introduces a multi-stage filtering strategy within the CFD loop, combining an absolute residual criterion, a residual reduction ratio, and a barycentric-map-based realizability constraint, enabling early identification and rejection of unstable and non-physical candidate models.

The RR-GEP framework achieves a 42.3\% reduction in computational cost relative to the B-GEP approach. Model rejection statistics show that a significant fraction of candidates are filtered out early, with rejection rates of 37.9\% (R-GEP) and 47.3\% (RR-GEP) in the first generation, and average rejection rates of 21.5\% and 37.8\%, respectively. The realizability filter is particularly effective, reducing non-realizable models at convergence from 58.4\% for B-GEP to 1.7\% in RR-GEP. 
The RR-GEP models exhibit bounded coefficient behavior, with magnitudes reducing from approximately $|\zeta_1| \approx 2.73$ (B-GEP) to $1.76$ (RR-GEP), leading to improved anisotropy characteristics and physically consistent stress representations. Trained on a canonical cylinder wake at $Re = 3900$, the models improve wake prediction and also reduce separation bubble length from $x/D \approx 4.5$ to $2.25$ and closely matches the LES reference dataset. 

Robust generalization is demonstrated across a rectangular cylinder, NACA0012 airfoil, and DARPA Suboff at significantly different Reynolds numbers (order magnitude $10^3$ to $10^6$) and flow regimes, while maintaining realizability.  Despite these substantial variations in geometry and flow physics, the learned models consistently maintain realizability across all cases and yield improved wake predictions, indicating strong transferability beyond the training configuration. This highlights the ability of the framework to discover closures that capture underlying turbulence physics rather than overfitting to a specific flow.
The framework enables efficient exploration of the model space while ensuring numerical robustness and physical fidelity, providing a strong foundation for next-generation turbulence model development. It can also be readily integrated with other machine learning approaches (e.g., \citep{McConkey2025Realisability}) to further reduce simulation time, offering a scalable pathway to significantly lower the computational cost of CFD-driven turbulence closure development.

\vspace{0.5mm}
\noindent \textbf{Acknowledgements}

Harshal Akolekar acknowledges the seed grant support from IIT Jodhpur (I/SEED/HDA/20230206).

\section*{Nomenclature}
\begin{supertabular}{ll}
$\bm a_{ij}$ & Anisotropy tensor \\
$c$ & NACA 0012 chord length \\
$C_i$ & Barycentric weights for limiting turbulence states\\
$C_n$ & Candidate models\\
$D$ & Circular cylinder diameter\\
$H$ & Rectangular cylinder height\\
$I_1, I_2$ & Invariants of strain-rate and rotation-rate tensors \\
$J$ & Cost function \\
$k$ & Turbulent kinetic energy \\
$L$ & Length of rectangular cylinder\\
$l_T$ & Turbulent length scale\\
$N_i$ & Iteration count \\
$P$ & Pressure \\
$P_k$ & Turbulent kinetic energy production \\
$Re$ & Reynolds number \\
$R^{(q)}$ & Residual of the $q^{th}$ variable  \\
$S$ & Length of DARPA Suboff  \\
$\bm S_{ij}$ & Mean strain-rate tensor \\
$\bm S'_{ij}$ & Deviatoric strain-rate tensor \\
$\bm T^{(n)}_{ij}$ & Tensor basis functions \\
$Tu_{\infty}$ & Inlet freestream turbulence intensity \\
$\bm U$ & Streamwise velocity \\
$\bm U_\infty$ &  Inlet free-stream velocity \\
$\bm u_i', \bm u_j'$ & Velocity fluctuations \\
$u'v'$ & Reynolds shear stress \\
$W$ & DARPA Suboff diameter \\
$x, y$ & Spatial coordinates \\
$x_b, y_b $ & Barycentric x and y coordinates \\
$y^+$ & Dimensionless wall distance\\
 \end{supertabular}
\vspace{0.1cm}

\noindent \textbf{Greek Symbols} 
\vspace{0.075cm}

\begin{supertabular}{ll}
$\alpha$ & Residual scaling parameter \\
$\Gamma$ & Residual reduction ratio between iterations $N_1$ and $N_2$ \\
$\gamma_{\min}$ & Minimum threshold for acceptable residual reduction \\
$\bm \delta_{ij}$ & Kronecker delta \\
$\epsilon_i$ & Residual threshold \\
$\zeta_n$ & Scalar coefficients of tensor basis functions \\
$\lambda_i$ & Eigenvalues of anisotropy tensor \\
$\nu_t$, $\mu_t$ & Eddy viscosity \\
$\bm \rho$ & Density\\
$\bm \tau$ & Turbulent time scale \\
$\bm \tau_{ij}$ & Reynolds stress tensor \\
$\bm \Omega_{ij}$ & Mean rotation-rate tensor \\
$ \omega$ & Specific dissipation rate \\

 \end{supertabular}

 \vspace{0.1cm}

\noindent \textbf{Abbreviations} 
\vspace{0.075cm}

\begin{supertabular}{ll}
CFD & Computational fluid dynamics \\
DARPA & Defense Advanced Research Projects Agency \\ 
EARSM & Explicit algebraic Reynolds stress model \\
GEP & Gene expression programming \\
LES & Large eddy simulation \\
NACA & National Advisory Committee for Aeronautics \\
RANS & Reynolds-averaged Navier–Stokes \\
R-GEP & Residual filtered GEP \\
RR-GEP & Realizability Residual-filtered GEP \\
B-GEP & Baseline GEP \\
SIMPLE & Semi-implicit method for pressure-linked equations\\

 \end{supertabular}

\section*{DATA AVAILABILITY}  
The data that support the findings of this study are available from the corresponding author upon reasonable request.

\printcredits

\bibliographystyle{model1-num-names}

\bibliography{cas-refs}

\end{document}